\documentclass[twoside,twocolumn,9pt]{article}
\usepackage{extsizes}
\usepackage[super,sort&compress,comma]{natbib} 
\usepackage[left=1.5cm, right=1.5cm, top=1.785cm, bottom=2.0cm]{geometry}
\usepackage{balance}
\usepackage{mathptmx}
\usepackage{sectsty}
\usepackage{graphicx} 
\usepackage{lastpage}
\usepackage[format=plain,justification=justified,singlelinecheck=false,font={stretch=1.125,small,sf},labelfont=bf,labelsep=space]{caption}
\usepackage{float}
\usepackage{fancyhdr}
\usepackage{fnpos}
\usepackage[english]{babel}
\addto{\captionsenglish}{%
  
}
\usepackage{array}
\usepackage{charter}
\usepackage[T1]{fontenc}
\usepackage[usenames,dvipsnames]{xcolor}
\usepackage{setspace}
\usepackage[compact]{titlesec}
\usepackage{hyperref}

\usepackage{epstopdf}

\hypersetup{pdfborder=0 0 0,colorlinks=true,citecolor=blue,linkcolor=blue}

\definecolor{cream}{RGB}{222,217,201}

\begin{document}

\pagestyle{fancy}
\thispagestyle{plain}
\fancypagestyle{plain}{
\renewcommand{\headrulewidth}{0pt}
}

\makeFNbottom
\makeatletter
\renewcommand\LARGE{\@setfontsize\LARGE{15pt}{17}}
\renewcommand\Large{\@setfontsize\Large{12pt}{14}}
\renewcommand\large{\@setfontsize\large{10pt}{12}}
\renewcommand\footnotesize{\@setfontsize\footnotesize{7pt}{10}}
\makeatother

\renewcommand{\thefootnote}{\fnsymbol{footnote}}
\renewcommand\footnoterule{\vspace*{1pt}%
\color{cream}\hrule width 3.5in height 0.4pt \color{black}\vspace*{5pt}} 
\setcounter{secnumdepth}{5}

\makeatletter 
\renewcommand\@biblabel[1]{#1}            
\renewcommand\@makefntext[1]%
{\noindent\makebox[0pt][r]{\@thefnmark\,}#1}
\makeatother 
\renewcommand{\figurename}{\small{Fig.}~}
\sectionfont{\sffamily\Large}
\subsectionfont{\normalsize}
\subsubsectionfont{\bf}
\setstretch{1.125} 
\setlength{\skip\footins}{0.8cm}
\setlength{\footnotesep}{0.25cm}
\setlength{\jot}{10pt}
\titlespacing*{\section}{0pt}{4pt}{4pt}
\titlespacing*{\subsection}{0pt}{15pt}{1pt}

\fancyhead{}
\renewcommand{\headrulewidth}{0pt} 
\renewcommand{\footrulewidth}{0pt}
\setlength{\arrayrulewidth}{1pt}
\setlength{\columnsep}{6.5mm}
\setlength\bibsep{1pt}

\makeatletter 
\newlength{\figrulesep} 
\setlength{\figrulesep}{0.5\textfloatsep} 

\newcommand{\topfigrule}{\vspace*{-1pt}%
\noindent{\color{cream}\rule[-\figrulesep]{\columnwidth}{1.5pt}} }

\newcommand{\botfigrule}{\vspace*{-2pt}%
\noindent{\color{cream}\rule[\figrulesep]{\columnwidth}{1.5pt}} }

\newcommand{\dblfigrule}{\vspace*{-1pt}%
\noindent{\color{cream}\rule[-\figrulesep]{\textwidth}{1.5pt}} }

\makeatother

\twocolumn[
  \begin{@twocolumnfalse}
\vspace{1em}
\sffamily


\noindent\LARGE{\textbf{Flattening conduction and valence bands for interlayer excitons in a moir\'e MoS$_2$/WSe$_2$ heterobilayer}}\\
\vspace{0.3cm} \\

\noindent\large{Sara Conti,$^{\ast}$\textit{$^{a}$} Andrey Chaves,\textit{$^{b}$} Tribhuwan Pandey,\textit{$^{a}$} Lucian Covaci,\textit{$^{a,c}$} Fran\c{c}ois M. Peeters,\textit{$^{a,b}$} David Neilson,\textit{$^{a}$} and Milorad V. Milo\v{s}evi\'c \textit{$^{a,c}$}} \\

\noindent\normalsize{
We explore the flatness of conduction and valence bands of interlayer excitons in MoS$_2$/WSe$_2$ van der Waals heterobilayers, tuned by interlayer twist angle, pressure, and external electric field. 
We employ an efficient continuum model where the moir\'e pattern from lattice mismatch and/or twisting is represented by an equivalent mesoscopic periodic potential.
We demonstrate that the mismatch moir\'e potential is too weak to produce significant flattening.  
Moreover, we draw attention to the fact that the quasi-particle effective masses around the $\Gamma$-point and the band flattening are \textit{reduced} with twisting.
As an alternative approach, we show (i) that reducing the interlayer distance by uniform vertical pressure can significantly increase the effective mass of the moir\'e hole, 
and (ii) that the moir\'e depth and its band flattening effects are strongly enhanced by accessible electric gating fields perpendicular to the heterobilayer, with resulting electron and hole effective masses increased by more than an order of magnitude leading to record-flat bands.
These findings impose boundaries on the commonly generalized benefits of moir\'e twistronics, while also revealing alternate feasible routes to achieve truly flat electron and hole bands to carry us to strongly correlated excitonic phenomena on demand.
} \\

 \end{@twocolumnfalse} \vspace{0.6cm}

  ]

\renewcommand*\rmdefault{bch}\normalfont\upshape
\rmfamily
\section*{}
\vspace{-1cm}

\footnotetext{\textit{$^{a}$~Department of Physics, University of Antwerp, Groenenborgerlaan 171, Antwerp 2020, Belgium,  E-mail: sara.conti@uantwerpen.be}}
\footnotetext{\textit{$^{b}$~Departamento de F\'{\i}sica, Universidade Federal do Cear\'a, Caixa Postal 6030, Fortaleza 60455-760, Brazil}}
\footnotetext{\textit{$^{c}$~NANOlab Center of Excellence, University of Antwerp, Antwerp 2020, Belgium}}




\section{Introduction}

The exciting report of superconductivity associated with flat bands in moir\'e twisted bilayer graphene \cite{Cao2018} opened the floodgates to inducing and tuning the strongly correlated states associated with flat bands by interlayer twisting.\cite{Yankowitz2019,torma2022} 
Flat bands and the associated correlated states should have a strong effect on the properties of interlayer excitons in bilayer systems, and in this paper we investigate these effects. 
The interlayer excitons are bound electron-hole pairs with the electrons and holes confined in spatially separated layers. 
Thanks to the separation, interlayer excitons have much longer lifetimes than intralayer excitons which are formed from electrons and holes in the same layer.\cite{Rivera2015}  
In this paper, we investigate the effect of band flattening on interlayer excitons.  

Twisted bilayer graphene is unsuitable for the purpose because of the need for an insulating barrier to separate the layers and confine the electrons and the holes to their respective layers. 
Such a barrier greatly reduces the effects of interlayer coupling, making the layers behave as if they were isolated.
The isotropic band structure of the low energy states of graphene is then not sensitive to an interlayer twist angle. 
Recent progress in the fabrication of high-quality van der Waals stacked bilayers has extended twisting to other 2D materials, most notably the transition-metal dichalcogenides (TMDs). \cite{scherer2021,kiese2022} 
A major advantage of TMDs here is that the two TMD layers can be different, forming a heterobilayer, and they can be chosen with a type-II interface.
A type-II interface confines the electrons and the holes in separate layers without need for an insulating spacer layer.
This allows formation of long-lived interlayer excitons in a moir\'e environment. \cite{Liu2022} 

Twisting in bilayers creates a periodic moir\'e pattern characterized by regions with different local stacking registries.
The moir\'e pattern yields potential landscapes for electrons and holes along the plane. \cite{Zhang2017a}
As a result, the properties of the excitons can be broadly tuned by flattening the electron and hole bands and changing their moir\'e potentials, leading to intriguing possibilities for novel technology devices. \cite{yu2017, Guo2020, Brzhezinskaya2021, Liu2022, Kononenko2022}

In homobilayers, moir\'e patterns occur only in twisted samples but in TMD heterobilayers, because of the incommensurability of the two different TMD lattices, moir\'e patterns are present even without twisting. 
Rotational alignment has been found to influence the interlayer coupling in homobilayers, \cite{vanderZande2014} but for TMD heterobilayers, the role and possible tunability of interlayer coupling remain open questions. 

In this paper, we investigate the electronic structure of electrons and holes in a moir\'e potential landscape for small interlayer twists, primarily in a MoS$_2$/WSe$_2$ van der Waals heterobilayer (vdWHB).  
We use an efficient continuum model parameterized from first principles.
Curiously, we find that the moir\'e bands for electrons and holes \textit{do not flatten with increasing twist angle}.  
On the contrary, the effective masses decrease as the twist angle grows. 

As an alternative flattening strategy, we investigate the effect of vertical pressure on the bands, directly linked with the reduction of the interlayer distance.
We find that pressure does enhance the strength of the moir\'e potential and results in larger effective masses.
As a further strategy, we also investigate the effect of an external electric field from vertical gating. 
We find that strong electric fields significantly enhance the moir\'e potential depths for the electrons and holes. 
This leads to increases in the effective masses up to two orders of magnitude, and the associated bands become ultra-flat.

\section{Results and discussion}

\begin{figure}[!ht]
\centering
\includegraphics[width=0.82\linewidth]{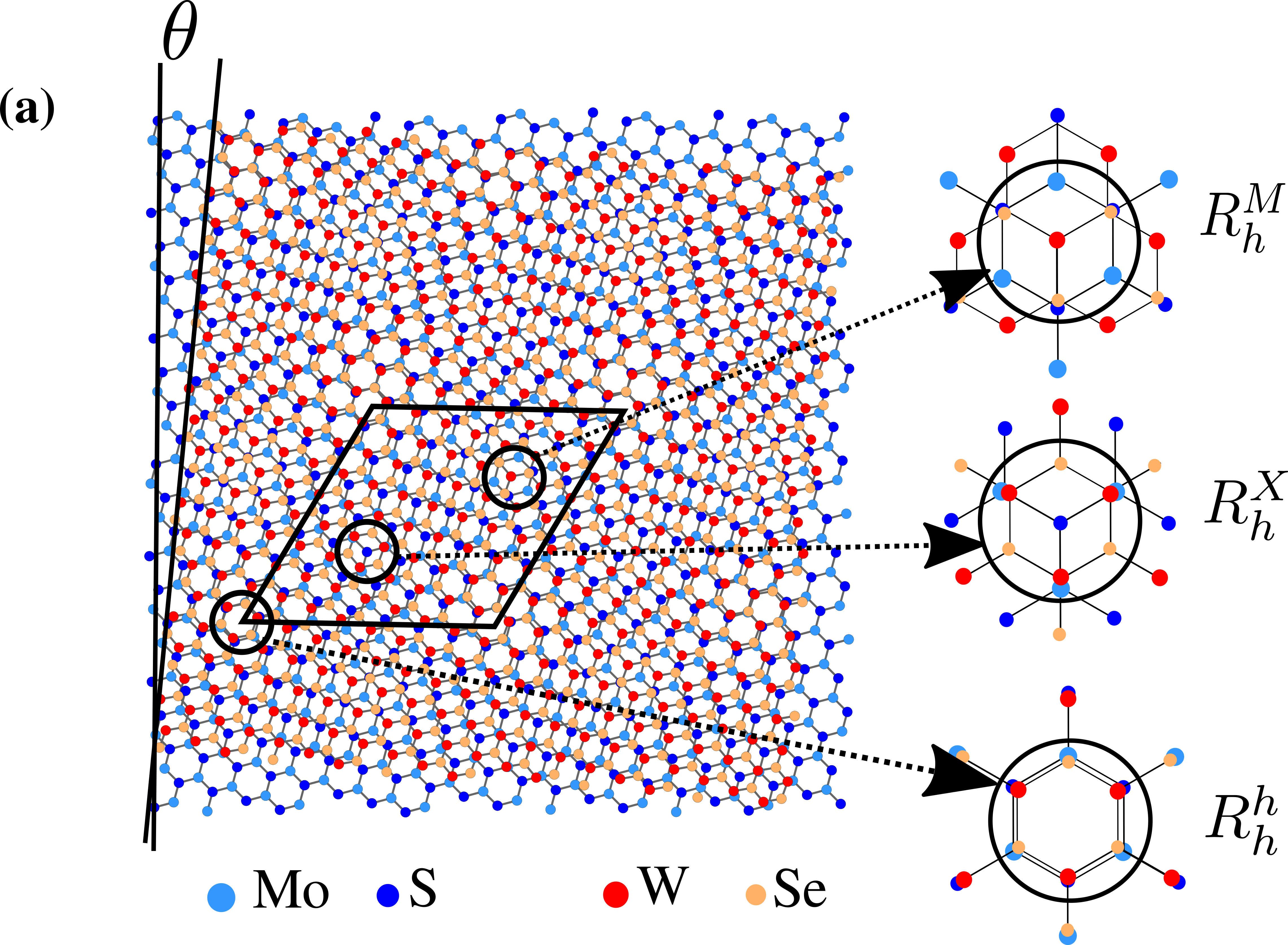}\\
\includegraphics[trim={0 0.4cm 0 0.4cm},clip,width=0.8\linewidth]{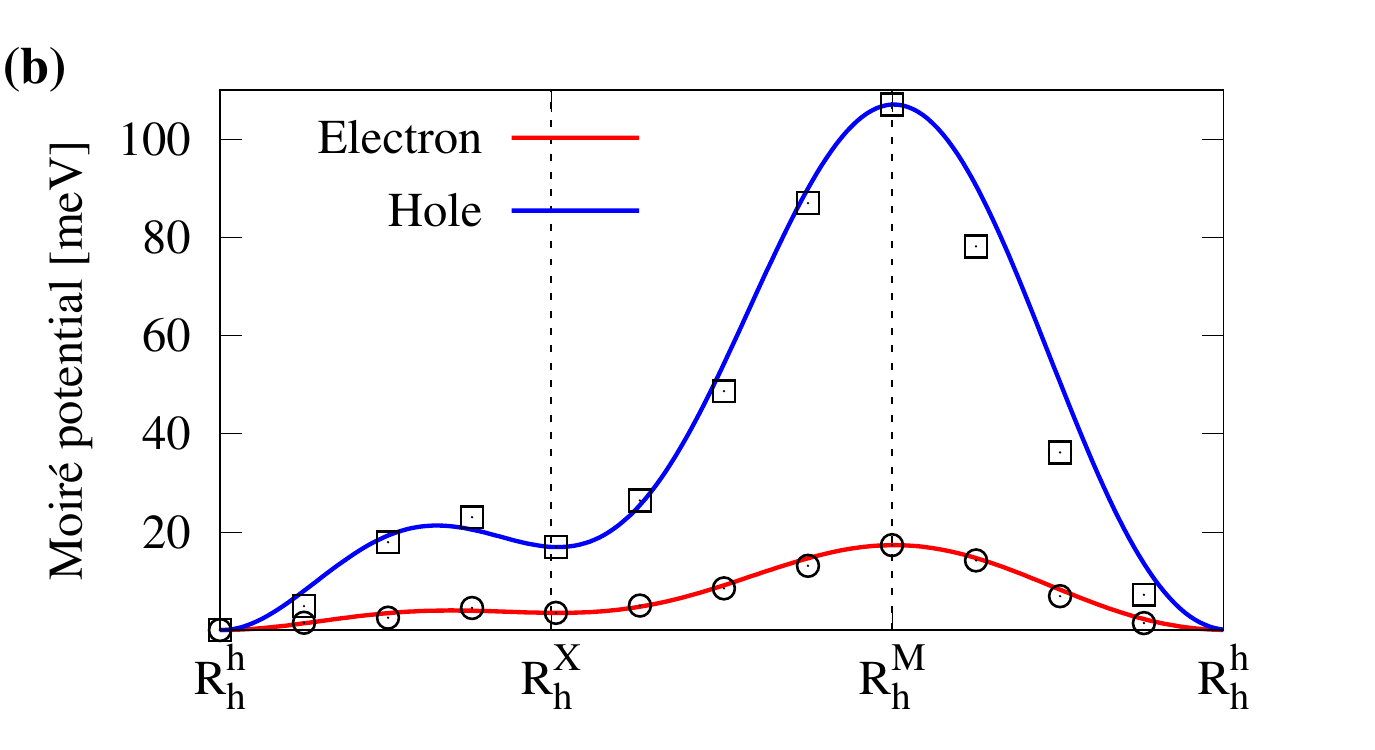}\\
\includegraphics[trim={0 0.1cm 0 0.4cm},clip,width=0.8\linewidth]{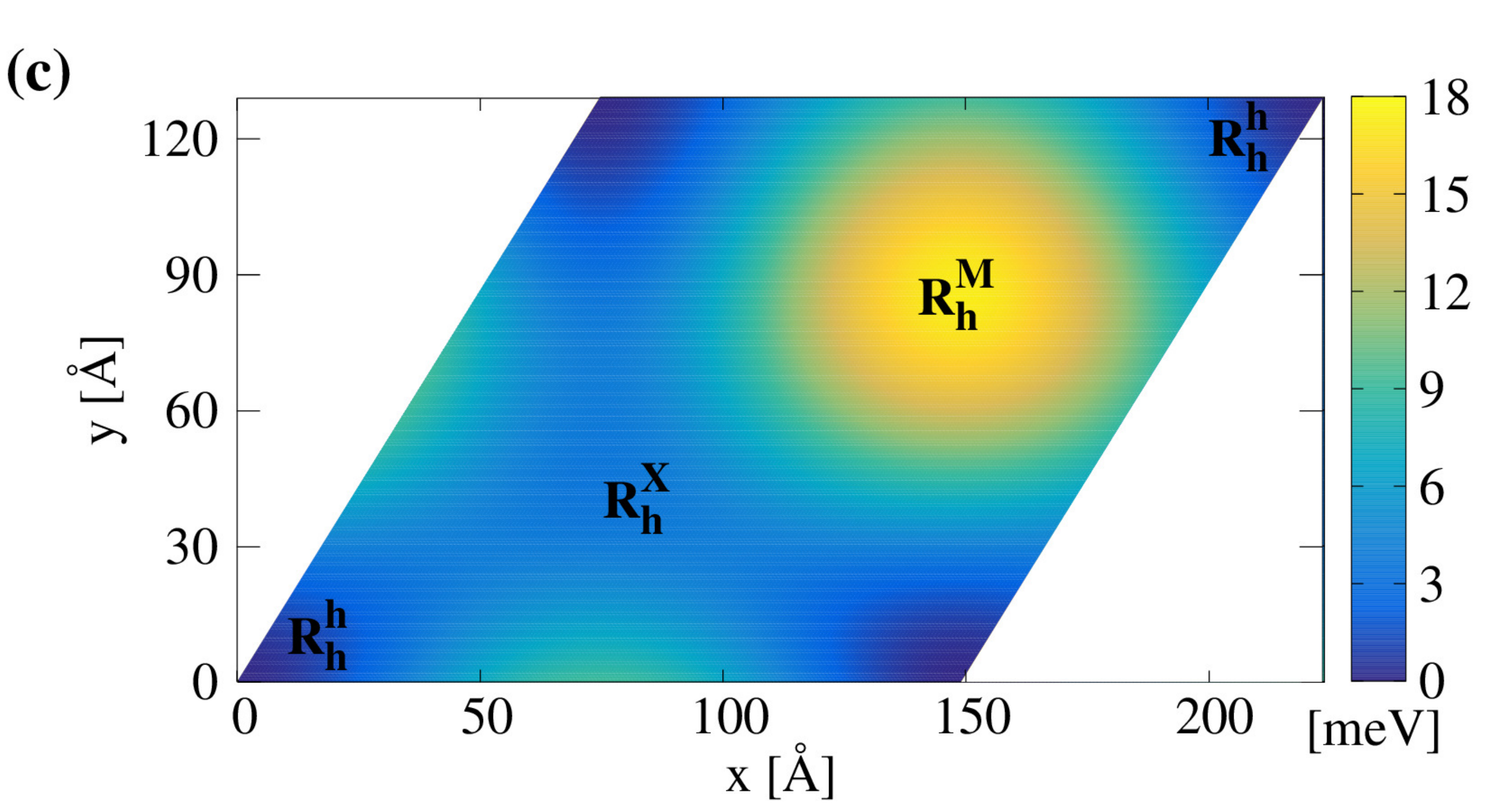}
\caption{(a) Sketch of a MoS$_2$/WSe$_2$ vdWHB with a small interlayer twist angle $\theta$ around zero, along with its moir\'e unit cell (black solid lines). Regions of local stacking registries $R_h^h$, $R_h^X$, and $R_h^M$  are identified. 
(b) Conduction and valence band edges at the $K$-point from the moir\'e potential of an infinite aligned MoS$_2$/WSe$_2$ vdWHB. Vertical dotted lines emphasize different stacking registers. 
(c) Moir\'e potential for the electrons obtained from the continuum model.}
\label{fig.sketch}
\end{figure}

We have used Density Functional Theory (DFT) to identify an optimal configuration of van der Waals heterobilayers. 
We searched for a heterobilayer with type-II band alignment and with the electrons and holes fully confined in their respective layers.  
We considered combinations of MoSe$_2$, MoS$_2$, WSe$_2$, WS$_2$, from which we selected type-II MoS$_2/$WSe$_2$ vdWHB with the electrons (holes) fully confined to the MoS$_2$ (WSe$_2$) layer. \cite{Zhang2017a}  
The band offsets are $\approx 220\ (640)$ meV. \cite{chiu2015,haastrup2018, gjerding2021}

Figure \ref{fig.sketch}a shows the crystal structure of a twisted MoS$_2$/WSe$_2$ vdWHB for a small angle $\theta$ centered on zero. 
The moir\'e unit cell is characterized by periodically alternating local stacking registries, identified as $R_h^h$, $R_h^M$, and $R_h^X$. 
Each stacking register exhibits different values for the energy of the conduction (valence) band minimum (maximum).

A further advantage of MoS$_2$/WSe$_2$ is that the other material combinations yield valence band maxima (VBM) which for some stacking registries lie at the $\Gamma$-point. 
In contrast, for MoS$_2$/WSe$_2$ the gap is always at the K-point regardless of the stacking registry, so only states at the K-point are involved. \cite{Kunstmann2018} 

Since the band edges at the K-point are mostly composed of d-orbitals of the metal atoms buried in between chalcogen atoms, the coupling between these band edge states in the two layers should not significantly change the interlayer band offsets.  
This allows us to consider the type-II interlayer exciton as the lowest energy case. 
Indeed, this is verified by DFT calculations of untwisted heterobilayers with different stacking orders, where band edges at the K-point are observed to be essentially superpositions of the monolayer bands. \cite{Chaves2018}

Figures \ref{fig.sketch}b and c show the effect of the moir\'e potential on the spatial evolution of the band edges for electrons and holes along the diagonal of the moir\'e unit cell in aligned MoS$_2$/WSe$_2$ vdWHB.
The band edges are obtained using a continuous model for the moir\'e potential (eqn (\ref{eq.V}) in Sec.~\ref{secCM}) with DFT parameters $V_{e,1} = -17.3$ meV, $V_{h,1} = -107.1$ meV, $V_{e,2} = -13.8$ meV, and $V_{h,2} = -90.2$ meV.
The energies are taken with respect to the minimum energy value corresponding to $R_h^h$ where for this particular vdWHB MoS$_2$/WSe$_2$ both the electrons and holes would be trapped.
The effect of the moir\'e pattern is much weaker for the electron bands than for the hole bands.
This is because the VBM charge density extends to both layers giving rise to non-zero interlayer mixing, while the conduction band minimum (CBM) charge density is completely confined to the MoS$_2$ layer (Fig.~\ref{fig.bandwf} in Sec.~\ref{secDFT}). Thus, the VBM is modulated by the interlayer moir\'e potential, while the CBM responds to the intralayer moir\'e potential. Because the interlayer moir\'e potential is much deeper than the intralayer potential, the changes in the VBM charge density are more pronounced than those in the CBM.

In the following subsections, the electrons (holes) are always confined in their MoS$_2$ (WSe$_2$) layer, which allows us to take the same effective masses $m_{e(h)} = 0.43\ (0.35) m_0$ throughout. \cite{haastrup2018,gjerding2021}

\subsection{Flattening the moir\'e band by twisting}\label{secflatT}
\begin{figure}[h]
\centering
\includegraphics[width =0.7\columnwidth]{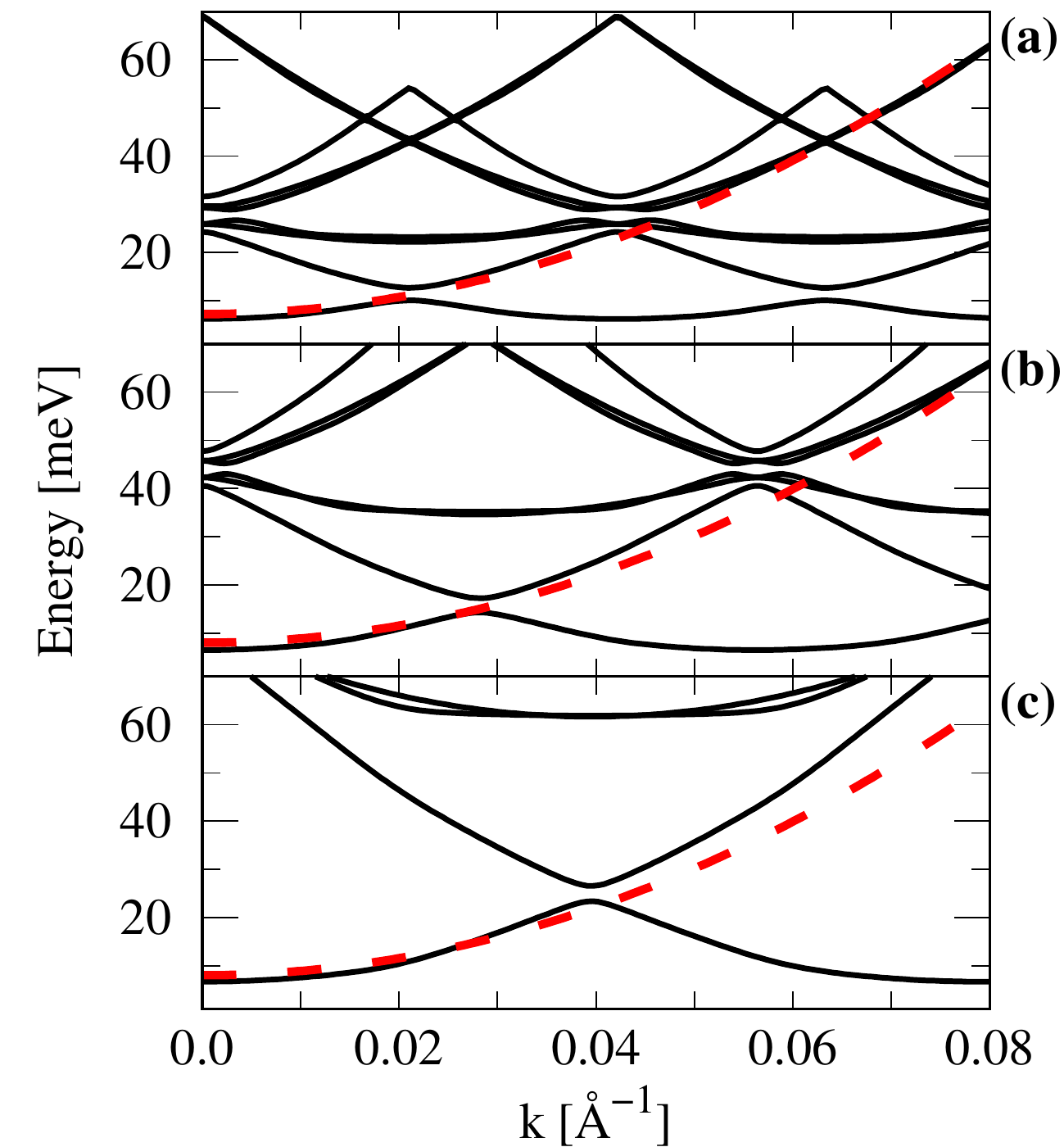}
\caption{Moir\'e bands for electrons in MoS$_2$/WSe$_2$ vdWHB with interlayer twist angle (a) $\theta =$ $0.5^\circ$, (b) $2.0^\circ$, and (c) $3.5^\circ$. Red dashed lines show for comparison a parabolic dispersion with electron effective mass $m_e = 0.43 m_0$ in an isolated MoS$_2$ layer.} 
\label{fig.moireelectron}
\end{figure}
\begin{figure}[h]
\centering
\includegraphics[width =0.7\columnwidth]{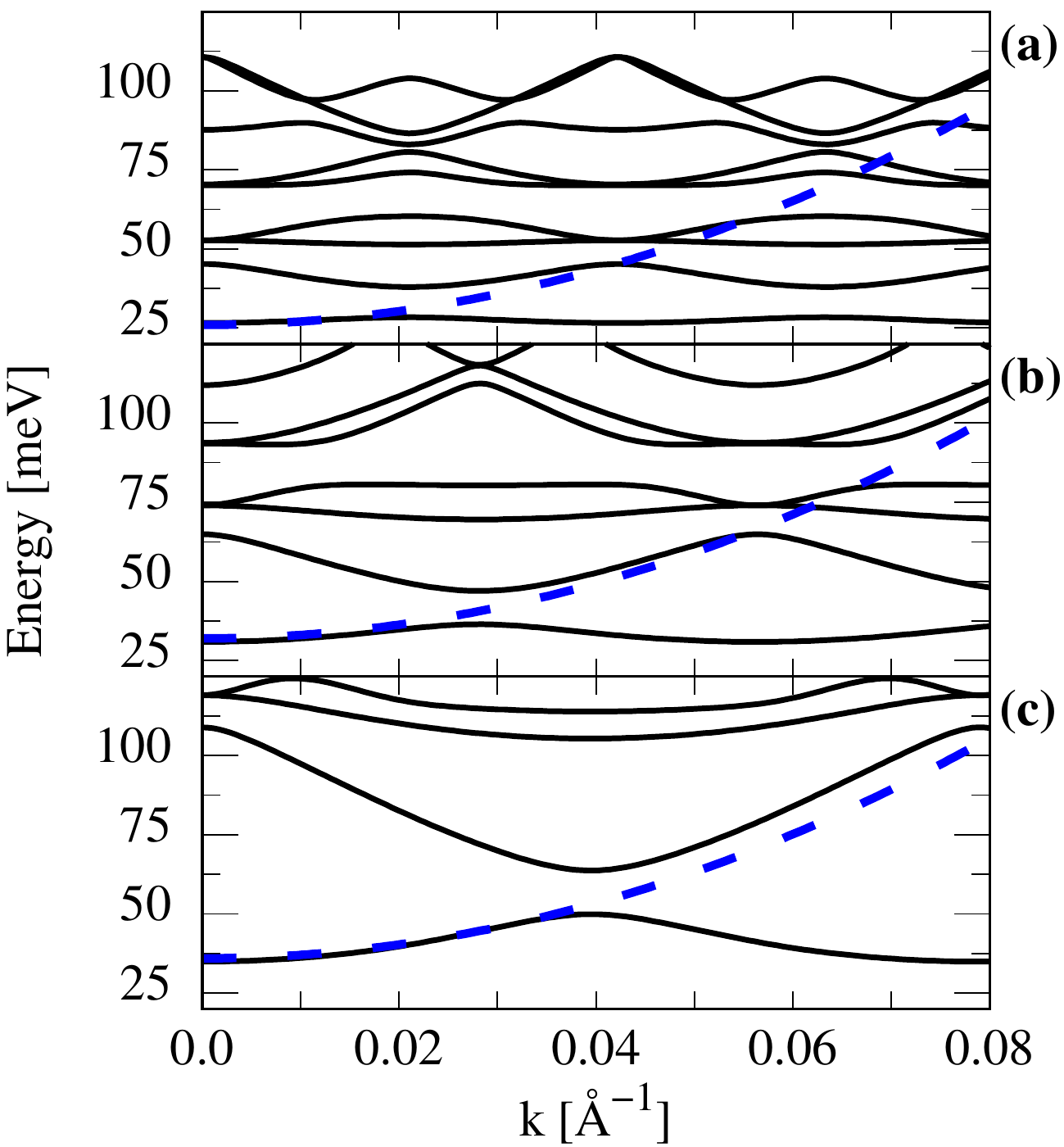}
\caption{Moir\'e bands for holes in MoS$_2$/WSe$_2$ vdWHB with interlayer twist angle (a) $\theta =$ $0.5^\circ$, (b) $2.0^\circ$, and (c) $3.5^\circ$. Blue dashed lines show for comparison a parabolic dispersion with hole effective mass $m_h = 0.35 m_0$ in an isolated WSe$_2$ layer.} 
\label{fig.moirehole}
\end{figure}

\begin{figure}[h]
\centering
\includegraphics[width=0.85\linewidth]{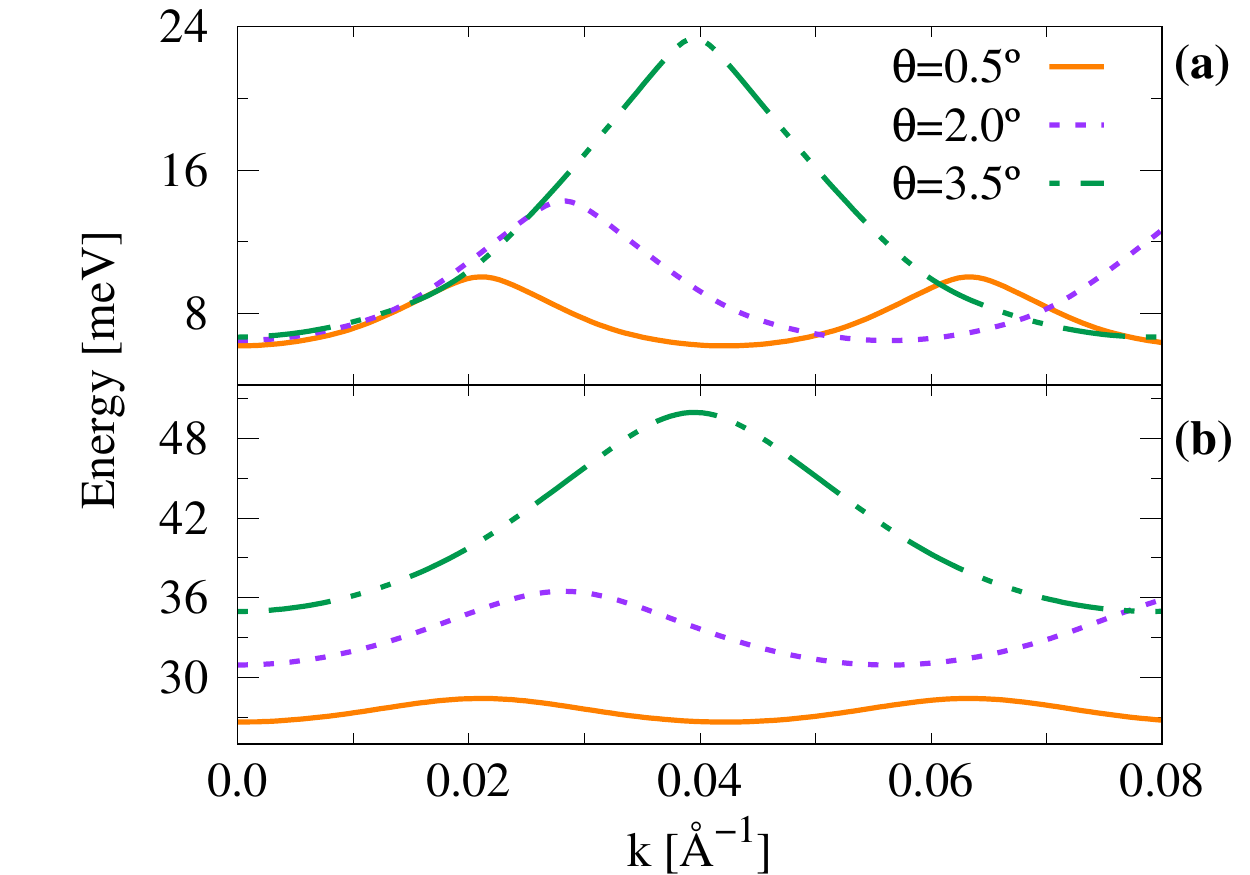}
\caption{Lowest energy moir\'e bands for (a) electrons and (b) holes in a MoS$_2$/WSe$_2$ vdWHB for different twist angles $\theta$.} 
\label{fig.lowenergymoire}
\end{figure}
\begin{figure}[h]
\centering
\includegraphics[width =0.75\linewidth]{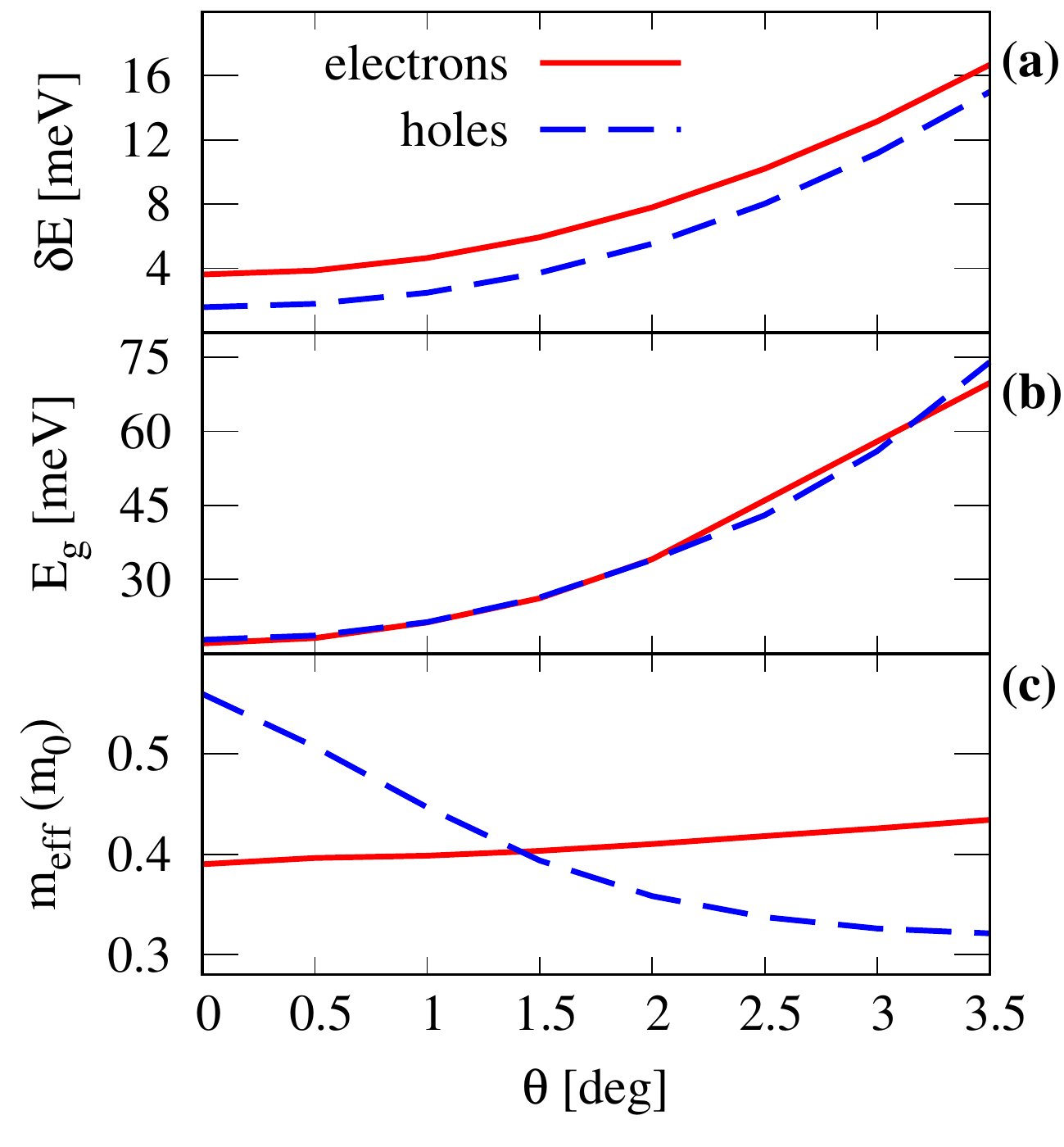}
\caption{(a) Ground-state bandwidths $\delta E$, (b) bandgap between the two lowest energy bands $E_g$, and (c) effective masses $m_\mathrm{eff}$ for electrons (red solid line) and holes (blue dashed line) in a MoS$_2$/WSe$_2$ vdWHB  as a function of the twist angle $\theta$.} 
\label{fig.flatteningTwist}
\end{figure}

Figures~\ref{fig.moireelectron} and \ref{fig.moirehole} show the band structure of moir\'e electrons and holes respectively, in the MoS$_2$/WSe$_2$ vdWHB along the $k_x$ direction.  
The interlayer twist angles are $\theta= 0.5^\circ$, $2^\circ$, and $3.5^\circ$. 
The $\Gamma$-point of the moir\'e Brillouin zone corresponds to the K-point of the Brillouin zone of the MoS$_2$ layer where the conduction band state is located. 
As the twist angle is increased, the period of the moir\'e pattern decreases, leading to larger and larger moir\'e Brillouin zones.

Figure~\ref{fig.lowenergymoire} compares the lowest energy conduction (a) and valence band (b) for different twist angles.
It is important to note that for small twist angles, bands may appear flat but this can be misleading. 
The apparent flatness is a result of a large anti-crossing at the moir\'e band edge combined with the short Brillouin zone length. 
Properties of compact moir\'e interlayer excitons involve states with wavelengths exceeding the abbreviated first Brillouin zone, requiring states from the higher Brillouin zones within the unfolded zone scheme. 
Figs.~\ref{fig.moireelectron} and \ref{fig.moirehole} show that the reconstructed unfolded bands follow closely the bands of electrons and holes in isolated layers, with their effective masses set at the K-point (red and blue dashed lines respectively).  
This means that these moir\'e quasi-particles  behave similarly to the original electrons and holes in isolated layers, with no significant increase in the quasi-particle effective masses. 

Figure~\ref{fig.flatteningTwist} shows the properties that are conventionally used to characterize the flattening of bands, namely the ground state bandwidths, the bandgap between the lowest energy bands, and the effective masses at the $\Gamma$-point. 
With increasing twist angle, the electron and hole bandwidths and bandgaps increase (Fig.~\ref{fig.flatteningTwist}a and b). 
Fig.~\ref{fig.flatteningTwist}c shows that the electron effective mass is insensitive to twisting, but that the hole effective mass significantly decreases with twisting.   
This difference in behavior reflects the shallowness of the electron moir\'e potential relative to the hole moir\'e potential (Fig.~\ref{fig.sketch}b). 

Band flatness is closely related to wave function localization. 
In terms of a tight-binding model of the moiré lattice, for a strongly localized wave function, the hopping between neighboring potential minima in the lattice will be small, resulting in flat bands.
The moir\'e potential landscape of untwisted MoS$_2$/WSe$_2$ vdWHB is too shallow to produce very flattened bands. 
Twisting this heterobilayer only further suppresses the original moir\'e potential because it decreases the distance between the moir\'e potential minima (larger moir\'e Brillouin zone), leading to stronger hopping energies between adjacent unit cells.  
Thus as shown in Fig.~\ref{fig.flatteningTwist}, twisting works against flattening of the bands in this heterobilayer, in contrast to twisted homobilayers.

To work towards band flattening with large effective masses and energy gaps, one must identify ways to deepen the confining moiré potential.
In what follows, we explore the deepening of moiré potentials through application of vertical pressure and electric fields.

\subsection{Flattening the moir\'e band by vertical pressure}\label{secflatP}

As an alternative strategy, we now look at the effect on the bands of pressure applied perpendicularly and uniformly across the layers to reduce the interlayer distance $d$. \cite{Ma2021}
Since the moir\'e potential is strongly affected by different interlayer couplings in the $R_h^h$, $R_h^M$, and $R_h^X$ stacking regions, any decrease $\delta d$ in the interlayer distance should efficiently control the depth of the moir\'e potential while leaving the moir\'e pattern unchanged.

The changes in the potential landscape due to the applied pressure are reflected in changes in the values of the parameters $V_{i,1}$ and $V_{i,2}$ in eqn (\ref{eq.V}).   
These determine the pressure-induced modifications on the band structures of the moir\'e electrons and holes in a MoS$_2$/WSe$_2$ vdWHB with the interlayer distance uniformly reduced in all regions by $\delta d$, up to $\delta d\sim 0.5$~\AA. 
When $\delta d \sim 0.4$ - $0.5$~\AA, for a few stackings the bandgap becomes indirect.
The value $\delta d=0.5$~\AA\ corresponds to an applied pressure $\sim 9$ GPa, well within the typical experimental range.\cite{xia2021}
Table~\ref{tab:1} lists the values of the parameters for different values of $\delta d$.  

\begin{table}[h]
\small
  \caption{\ Parameters (in meV) for reconstructing moiré potentials for electrons and holes [eqn (\ref{eq.V}) in Sec.~\ref{secCM}] under applied vertical pressure.}
  \label{tab:1}
  \begin{tabular*}{0.48\textwidth}{@{\extracolsep{\fill}}lllll}
    \hline
$\delta d$ (\AA\,) & $V_{e, 1}$ & $V_{e, 2}$ & $V_{h, 1}$ & $V_{h, 2}$ \\
    \hline
0.0 & -17.3 & -13.8 & -107.1 & -90.2\\
0.2  &  -19.1 &   -19.6 &  -138.1 &  -137.2 \\
0.4  &  -22.0 &   -24.4 &  -168.9 &  -199.2 \\
0.5 &  -24.5 &   -28.5 &  -182.8 & -237.4 \\
    \hline
  \end{tabular*}
\end{table}

\begin{figure}[h]
\centerline{\includegraphics[width = 0.8\linewidth]{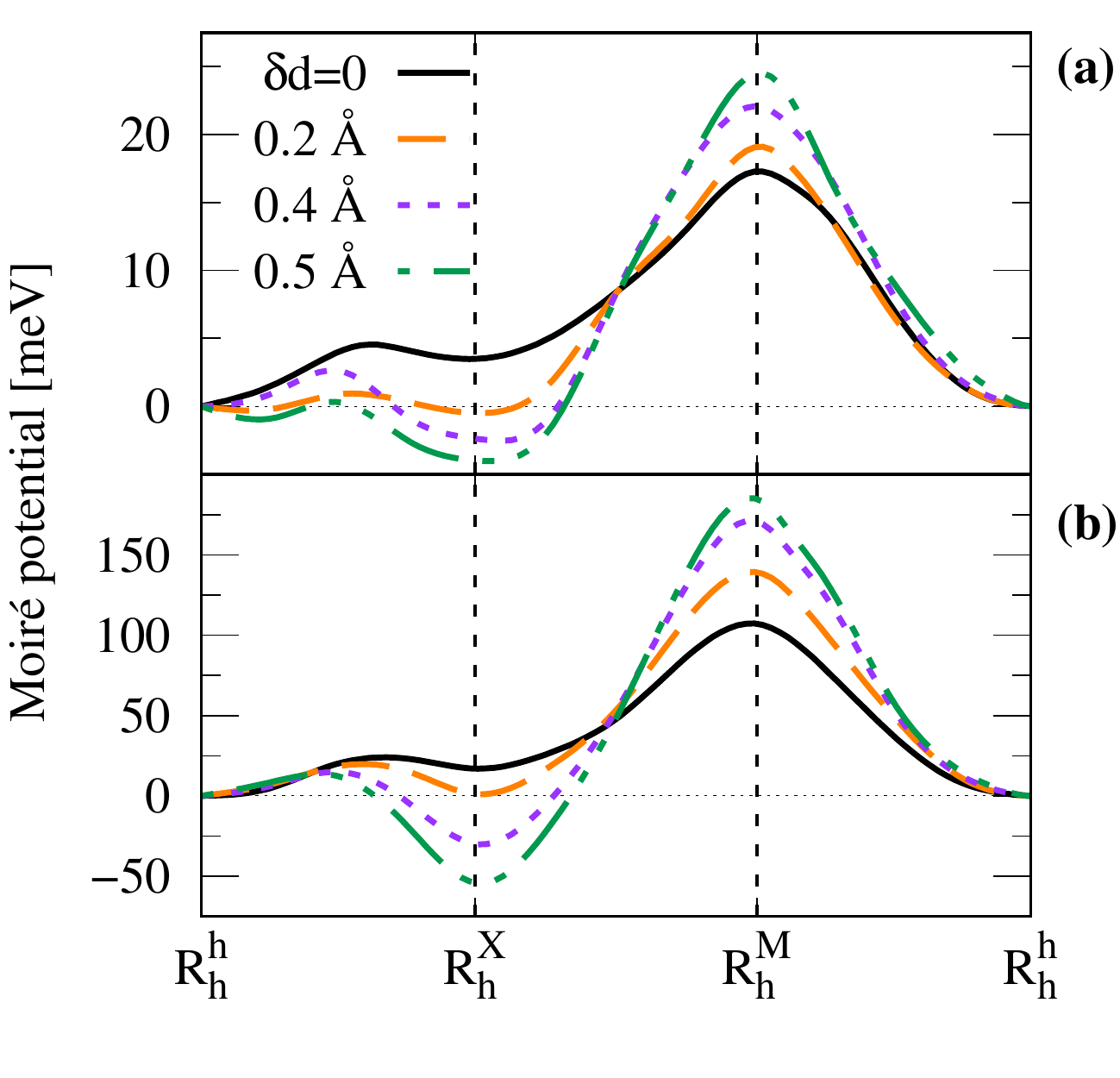}}
\caption{(a) Conduction and (b) valence band edges at the K-point from the moir\'e potential of a MoS$_2$/WSe$_2$ vdWHB with $\theta=0.5^\circ$, for decreases in the interlayer spacing $\delta d = 0$ (black solid line), $0.2$ \AA\ (orange dashed), $0.4$ \AA\ (purple dotted), and $0.5$ \AA\ (green dash-dotted line).} 
\label{fig.moirewithstrain}
\end{figure}

Figure~\ref{fig.moirewithstrain} shows the conduction band (a) and valence band (b) edges along the diagonal of the moir\'e unit cell of MoS$_2$/WSe$_2$ vdWHB as a function of $\delta d$. 
The applied pressure enhances the interlayer coupling and this increases the depth of the moir\'e potential for $R_h^X$ stacking.  
At $\delta d\sim 0.2$~\AA, the lowest energy state switches from the $R_h^h$ to the $R_h^X$ stacking.  

\begin{figure}[h]
\centering
\includegraphics[width = 0.85\linewidth]{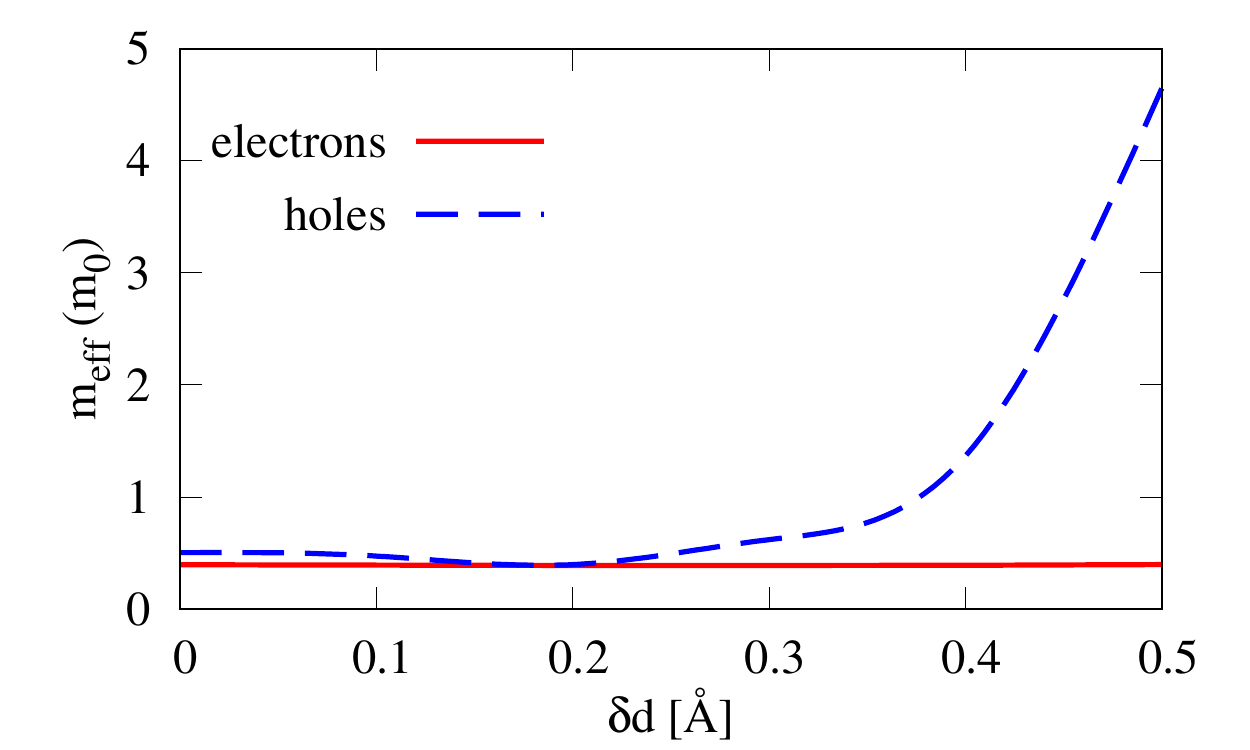}
\caption{Electron and hole effective masses $m_\mathrm{eff}$ in a MoS$_2$/WSe$_2$ vdWHB with $\theta=0.5^\circ$, as a function of the decrease $\delta d$ in the interlayer spacing  from vertical pressure.} 
\label{fig.masseswithstrain}
\end{figure}

We see from Fig.~\ref{fig.masseswithstrain} that, because of the switch, reducing the interlayer spacing has little effect on $m_\mathrm{eff}$ until $\delta d\sim0.2$~\AA .
When $\delta d>0.2$~\AA, the hole moir\'e potential at $R_h^X$ deepens to such an extent that, by $\delta d= 0.5$~\AA, the moir\'e hole effective mass has increased by nearly an order of magnitude compared with its value in an isolated WSe$_2$ monolayer. 

This is because the interlayer mixing of the VBM charge density significantly increases with decreasing interlayer distance.  
However, Fig.~\ref{fig.moirewithstrain}a shows for electrons that the variations in the moir\'e potential are small, and we see that the moir\'e electron effective mass does not change significantly with $\delta d$.   
Indeed, the CBM charge density is not expected to be sensitive to $\delta d$, since the conduction band states are strongly localized in the Mo atoms, buried in between the S atoms, which prevents electron states from undergoing significant modifications in the charge density due to the presence of another layer (Fig.~\ref{fig.bandwf}a in Sec.~\ref{secDFT}).
Valence band states, on the other hand, have their charge that also extends among the Se atoms, making them more susceptible to different interlayer distances and even to the stacking order (see Fig.~\ref{fig.bandwf}b in Sec.~\ref{secDFT}). 

We conclude that with applied uniform vertical pressure one obtains significant band flattening and an increase in the effective mass, but only for the valence band states.

\subsection{Flattening the moir\'e band by vertical gating}\label{secflatE}

\begin{figure}[h]
\centerline{\includegraphics[width =0.75\linewidth]{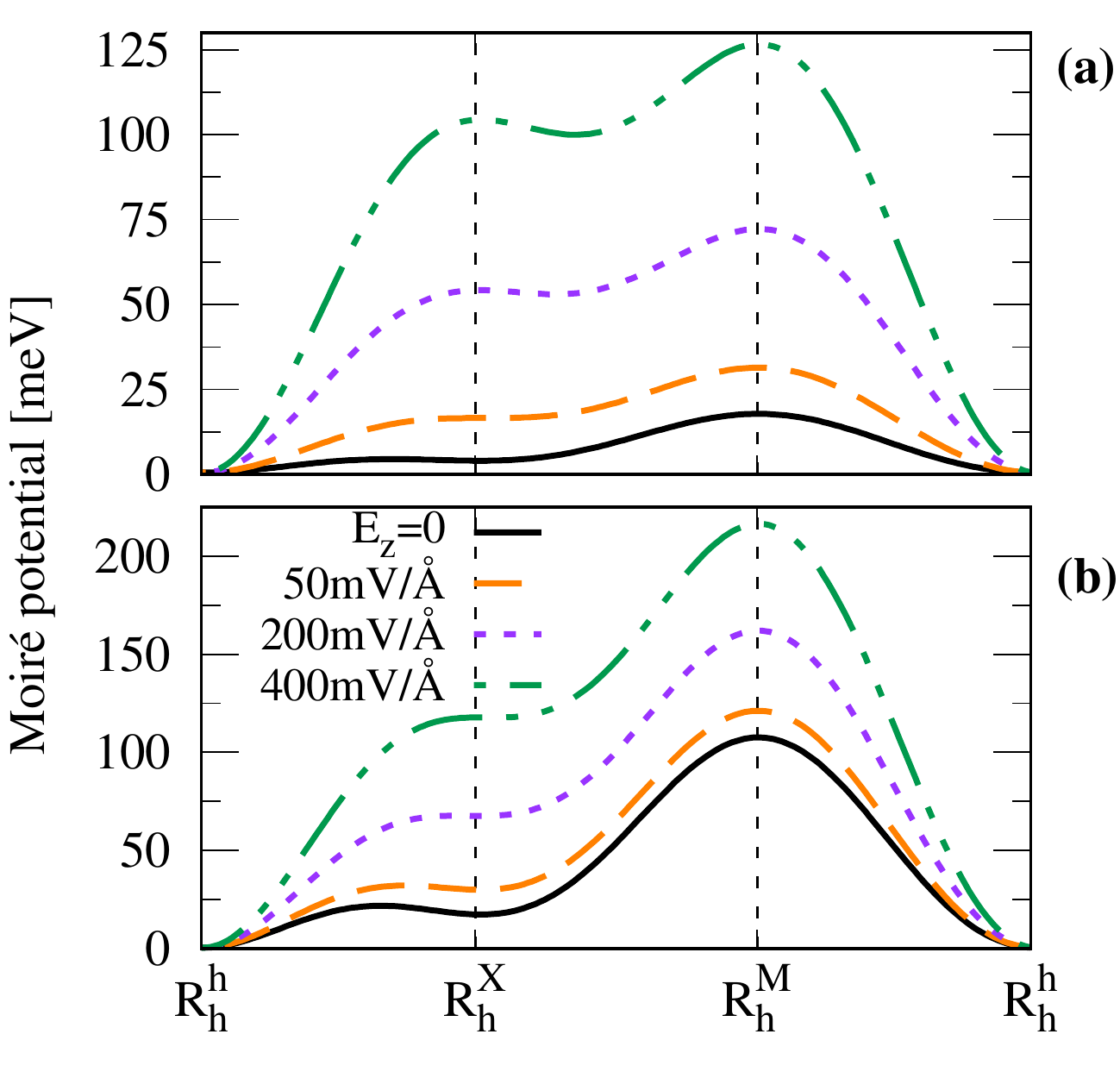}}
\caption{Conduction and (b) valence band edges at the K-point from the moir\'e potential of MoS$_2$/WSe$_2$ vdWHB with $\theta = 0.5^\circ$, for unbiased case (black solid line), and under applied perpendicular electric fields $E_z$ of $50$ mV/\AA\, (orange dashed), $200$ mV/\AA\, (purple dotted), and $400$ mV/\AA\, (green dash-dotted line).} 
\label{fig.moirepotentials}
\end{figure}

\begin{figure}[h]
\centerline{\includegraphics[width = 0.8\linewidth]{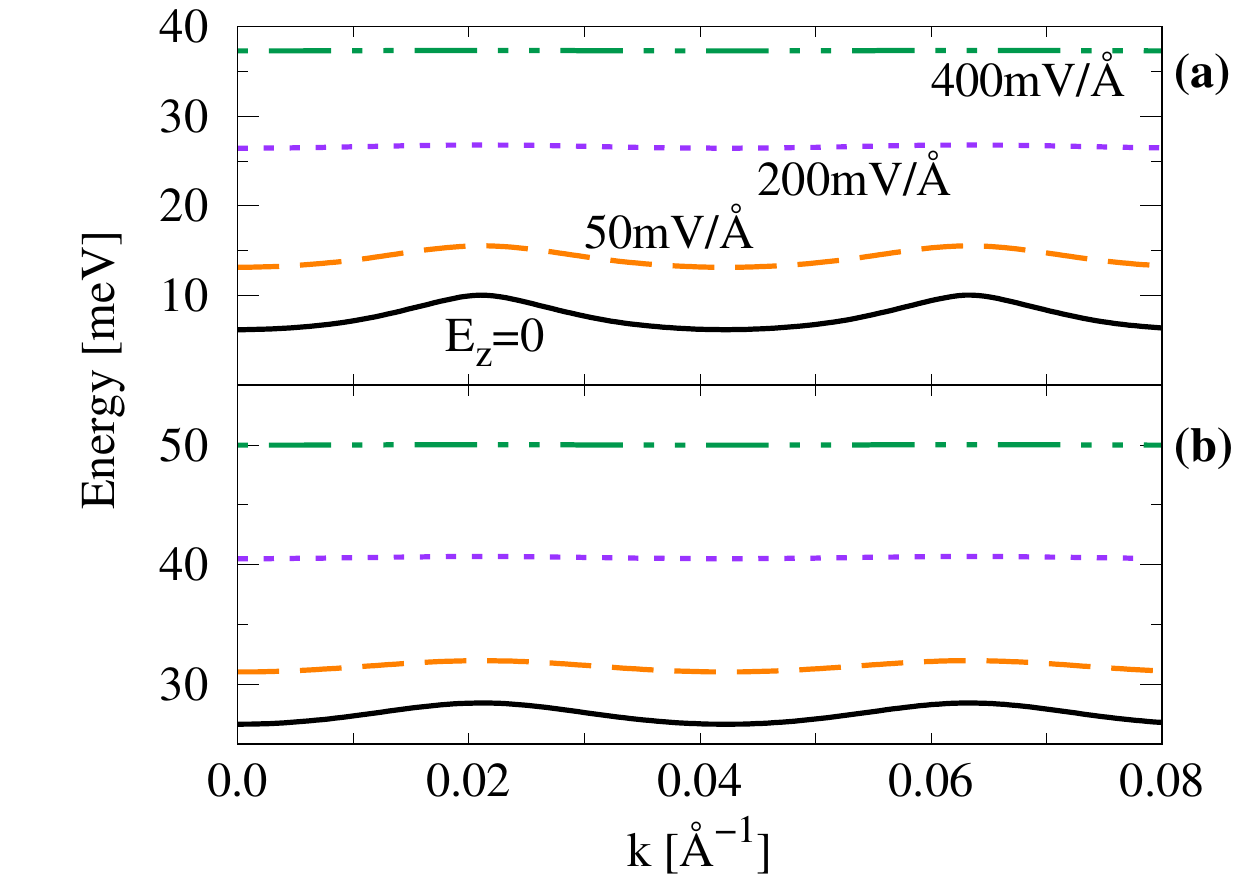}}
\caption{Lowest energy moir\'e bands for (a) electrons and (b) holes in a MoS$_2$/WSe$_2$ vdWHB with $\theta = 0.5^\circ$, for unbiased case (black solid line), and under applied perpendicular electric fields $E_z$ of $50$ mV/\AA\, (orange dashed), $200$ mV/\AA\, (purple dotted), and $400$ mV/\AA\, (green dash-dotted line).} 
\label{fig.lowenergyfield}
\end{figure}
\begin{figure}[h]
\centering
\includegraphics[width =0.7\columnwidth]{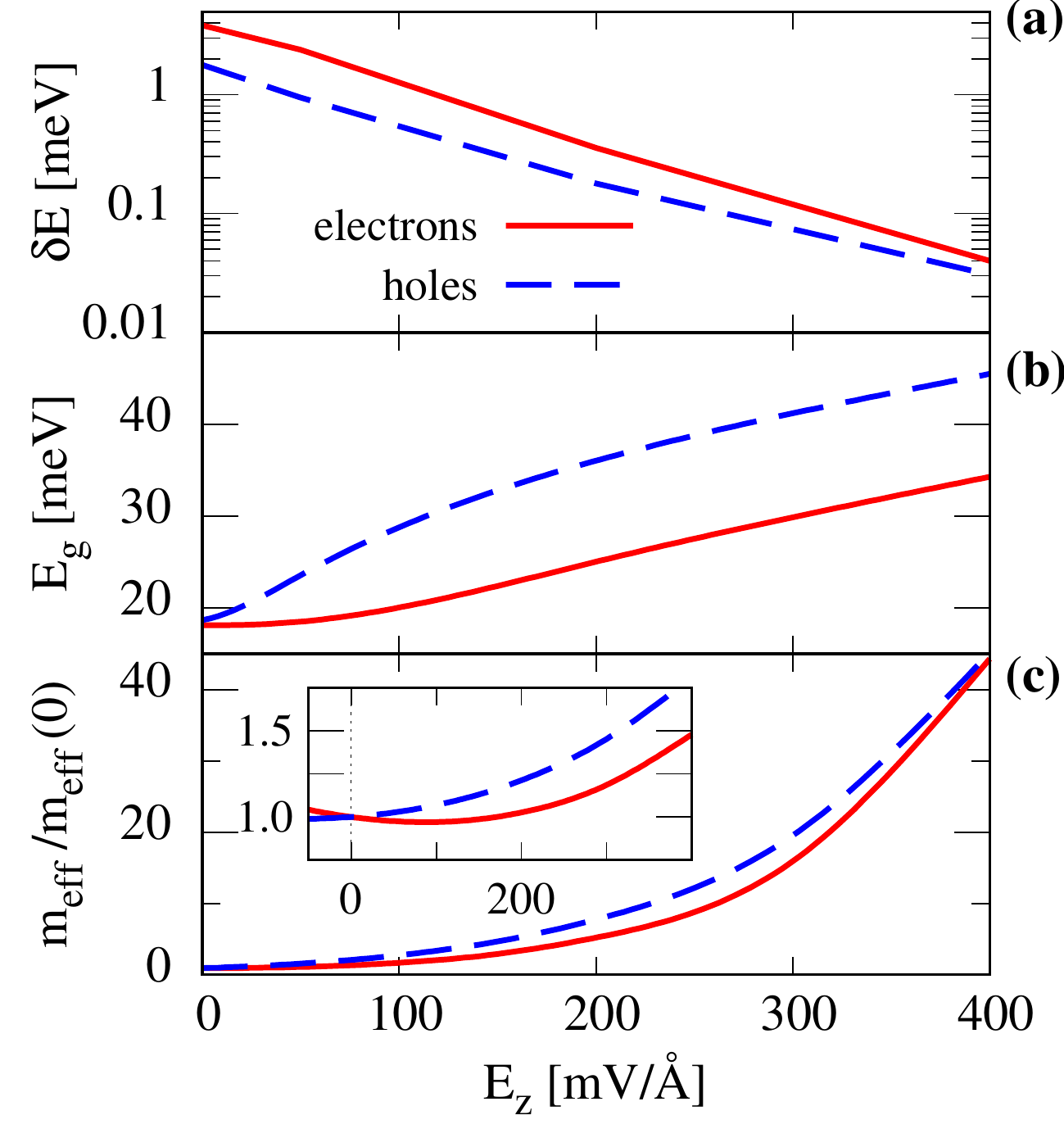}
\caption{  (a) Ground-state bandwidths $\delta E$, (b) bandgap between the two lowest energy bands $E_g$, and (c) effective masses $m_\mathrm{eff}$ for electrons (red solid line) and holes (blue dashed line) in a MoS$_2$/WSe$_2$ vdWHB  with $\theta=0.5^\circ$ as a function of the applied perpendicular electric field $E_z$. 
For clarity, $m_\mathrm{eff}$ is scaled to the values for zero field $m_\mathrm{eff}(E_z=0)$. 
Inset in (c): $m_\mathrm{eff}/m_\mathrm{eff}(0)$ for $\theta = 3.5^{\circ}$.} 
\label{fig.flatteningField}
\end{figure}

As a final strategy, we investigate the effect on the bands of an external perpendicular electric field from gating. 
The electric dipole of interlayer excitons couples with the electric field $E_z$, and this will affect the moir\'e potential. 
From the Stark effect, the electric field $E_z$ shifts the CBM and VBM energy levels in MoS$_2$ and WSe$_2$.
An electric field applied from the MoS$_2$ layer to the WSe$_2$ layer should decrease the band gap. 
A field in the opposite direction would increase the gap, and  for  relatively small fields the electrons and/or holes can change layers. \cite{Guo2020}   
The depth of the moir\'e potential landscape and the moir\'e effective masses can be readily increased and tuned by a field pointing from the MoS$_2$ to the WSe$_2$, since the $R_h^h$ stacking region containing the minimum of the moir\'e potential has the largest interlayer distance, \cite{Zhang2017a,yu2017} and hence the largest dipole moment.

Figure~\ref{fig.moirepotentials} shows the conduction and valence band edges of the MoS$_2$/WSe$_2$ vdWHB for different applied uniform electric fields. 
We find that the moir\'e potential indeed becomes progressively deeper with increasing $E_z$, while the $R_h^h$ always remains the stacking with minimum energy. 
The deepening of the moir\'e potential is much larger than for the case with pressure (Fig.~\ref{fig.moirewithstrain}).

Figure~\ref{fig.lowenergyfield} shows that with applied electric field, the lowest energy conduction band (a) and valence band (b) become remarkably flat.  
Fig.~\ref{fig.flatteningField}a shows bandwidths $\delta E$ as small as $\sim10^{-2}$ meV, the narrowest reported to date.
Fig.~\ref{fig.flatteningField}b shows the corresponding energy bands $E_g$ which increase by a factor of two, and Fig.~\ref{fig.flatteningField}c shows that there is a dramatic increase in both the electron and hole effective masses for vertical electric fields $E_z>100$ 
mV/\AA.  
The dramatic increase in the masses is severely curtailed when the twist angles are increased (inset Fig.~\ref{fig.flatteningField}), demonstrating once again the detrimental effects of twisting in this heterobilayer.

This is an exciting result, pointing to the possibility of achieving ultra-flat bands with associated strongly correlated interlayer exciton states in MoS$_2$/WSe$_2$ vdWHB by means of an applied electric field, without need for twisting.

\subsection{Effect of reconstruction of moiré lattice on band flattening}
\begin{figure*}[h]
\centering
\includegraphics[trim={1.4cm 0.cm 2.5cm 0.4cm},clip,width=0.7\linewidth]{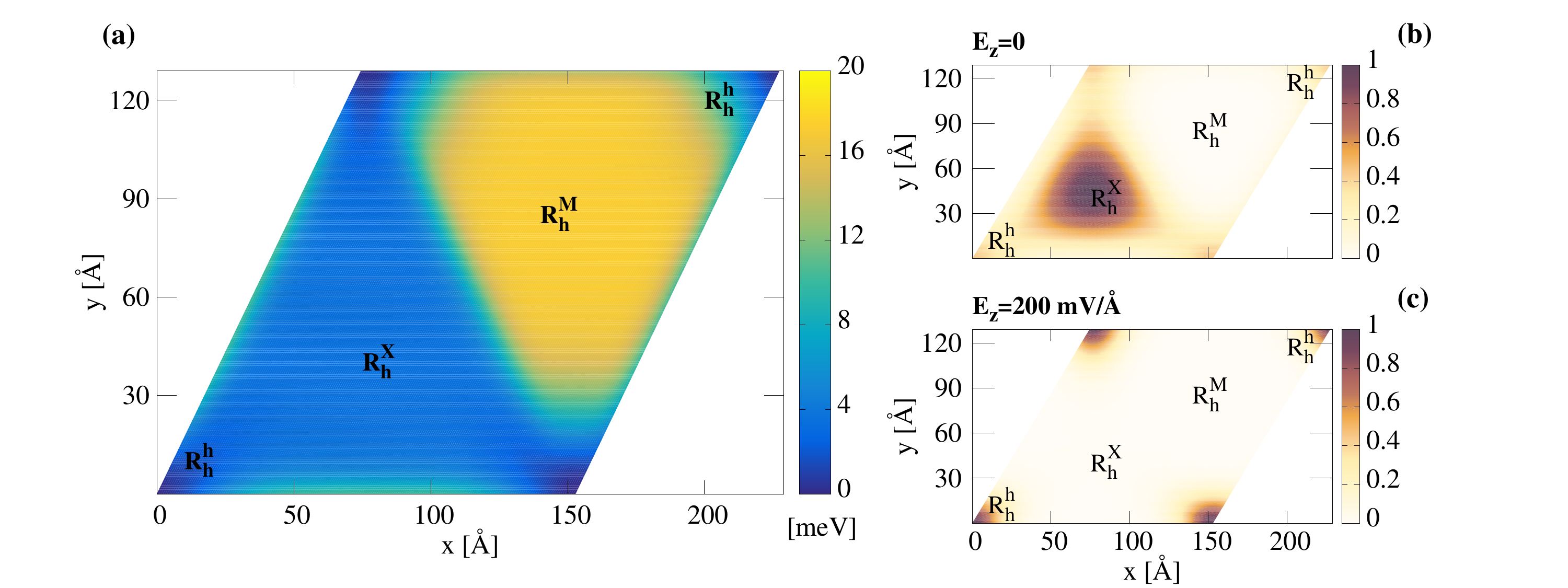}
\caption{(a) Moir\'e potential for the electrons in the reconstructed lattice of a MoS$_2$/WSe$_2$ vdWHB with sufficiently small twist angle.  
The potential in the vicinity of the confined $R_h^h$ region is modeled by a Gaussian function of width $d$ = 15~\AA .\cite{rosenberger2020twist}
Spatial distribution of the electron probability density for the reconstructed lattice, (b) in the absence of applied electric field, and (c) with perpendicular electric field 200 mV/\AA\,.}
\label{fig:reconel}
\end{figure*}

Recent experiments have reported lattice reconstruction upon relaxation of MoSe$_2$/WSe$_2$ heterobilayers with near-zero twist angles $\theta<1^{\circ}$. \cite{rosenberger2020twist, he2021moire}
In this combination of materials, the formation energies of the $R_h^M$ and $R_h^X$ stacking registries are almost degenerate and significantly smaller than the $R_h^h$ formation energy. 
Consequently, the regions of the original moiré pattern with these two stacking configurations expand and undergo reconstruction.
In our MoS$_2$/WSe$_2$ heterobilayers, the formation energies exhibit the same features as MoSe$_2$/WSe$_2$ (Fig.~\ref{energy_all}a in Sec.~\ref{secDFT}). 
Therefore, we expect the lattice reconstruction at small twist angles would be similar.

As a consequence, the pattern of the electron and hole effective potentials would no longer be the original moiré pattern, depicted in Fig.~\ref{fig.sketch}c for electrons, but rather it would acquire a different form, as shown in Fig.~\ref{fig:reconel}a.
The result would be a super-lattice composed of triangular domains with $R_h^M$ or $R_h^X$ registries with a strongly-reduced $R_h^h$ region connecting the corners of the triangles.
We assume there is no change in lattice parameters upon reconstruction, so the potentials in these regions will be nearly the same as shown in Fig.~\ref{fig.sketch}b.

\begin{figure}[h]
\centering
\includegraphics[width =0.7\columnwidth]{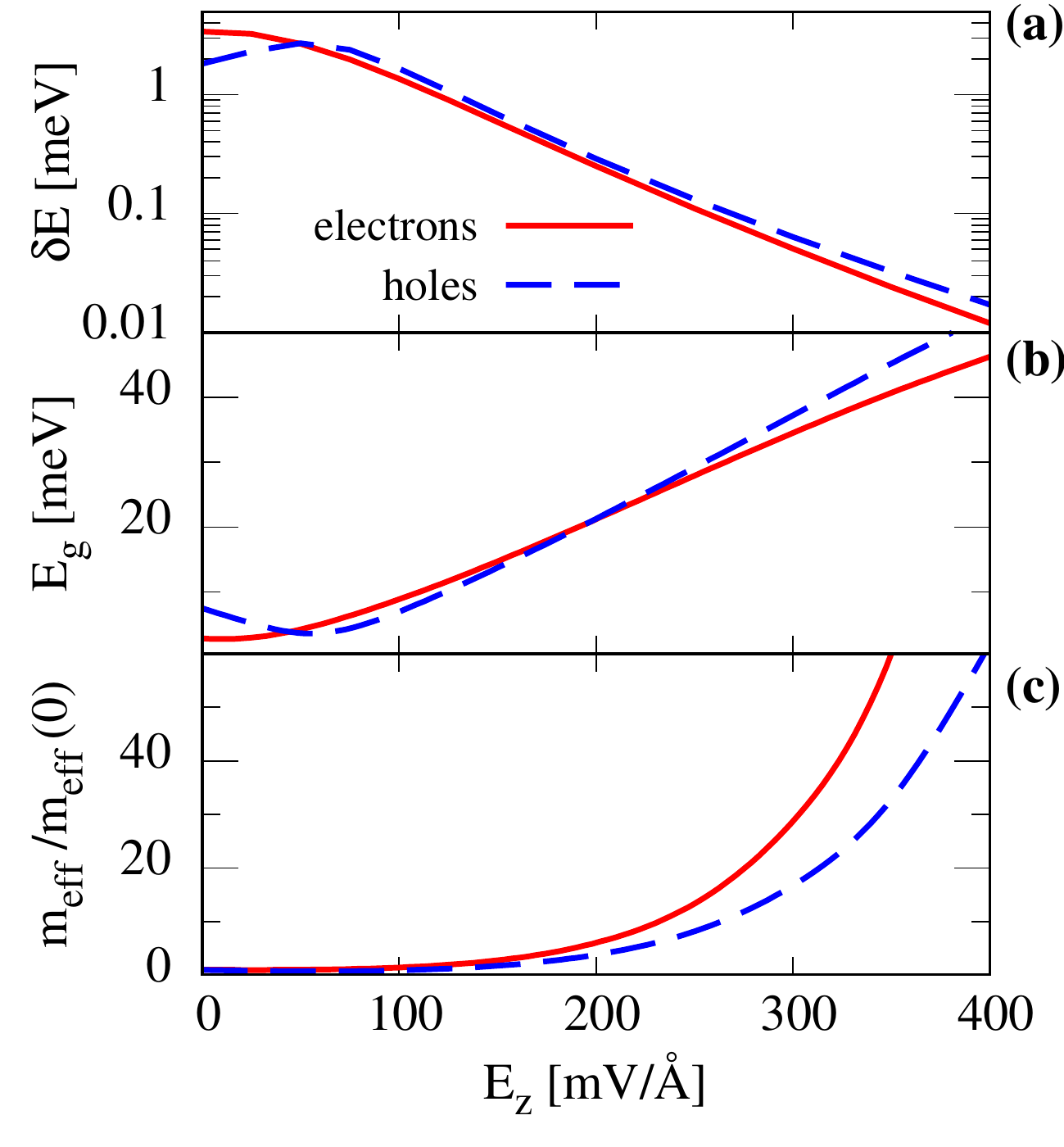}
\caption{After small angle lattice reconstruction in a MoS$_2$/WSe$_2$ vdWHB, (a) ground-state bandwidths $\delta E$, (b) bandgap between the two lowest energy bands $E_g$, and (c) effective masses $m_\mathrm{eff}$ for electrons (red solid line) and holes (blue dashed line) as a function of the perpendicular electric field $E_z$.
$m_\mathrm{eff}(0)$ is for zero electric field.}
\label{fig.flatteningField_REC}
\end{figure}

Figure \ref{fig.flatteningField_REC} shows, for electrons (red solid lines) and holes (blue dashed lines), the bandwidths $\delta E$ (a), the corresponding band gaps $E_g$ between the ground and first excited states (b), and the increase in both the electron and hole effective masses (c) as a function of the perpendicular electric field. 
At zero electric field, the narrow and shallow potential in the $R_h^h$ region is not able to confine electron and hole wave functions.
They are strongly localized but within the triangular $R_h^X$ patches (Fig.~\ref{fig:reconel}b).
They exhibit small overlap with adjacent triangles, and consequently, the hopping parameters are small compared with the  unreconstructed moiré potential.
This leads to higher electron and hole effective masses in the reconstructed case.

In contrast to the unreconstructed case, here a weak electric field decreases very slightly the effective masses and the band flattening. 
This is because a weak electric field deepens the confinement potential in the $R_h^h$ region, linking wave functions in the triangular $R_h^X$ regions and thus increasing the overlap between wave functions in neighboring triangles.
This leads to higher hopping parameters and hence less band flattening. 
However, except for weak fields, the deeper $R_h^h$ potential will fully confine the electron and hole wave functions.
The transition from wave functions confined in the triangular $R_h^X$ regions at zero field to wave functions strongly confined in the $R_h^h$ region is verified in the contourplots of the electron probability densities for zero and 200 mV/\AA\, electric fields (Figs.~\ref{fig:reconel}b-c). 
This confinement again leads to a decrease in the hopping parameters and a significant flattening of the bands, just as in the unreconstructed case.

Figure \ref{fig.flatteningField_REC}c shows that for electrons (holes), the electric field still strongly increases the effective masses by a factor 60,  decreases the bandwidths $\delta E$ by 99.65\% (99.86\%), and increases the gaps $E_g$ by a factor 10.
We see that the dramatic increase of the effective masses with electric field and the associated sub-meV bandwidths are very similar to that for the unreconstructed moiré lattices reported in Sec.~\ref{secflatE}.
Therefore the conclusions in the main manuscript about band flattening using applied electric fields remain equally valid in the presence of reconstruction. 

\subsection{Properties of interlayer excitons in moir\'e potential under band-flattening}\label{secProp}

Using the single-particle energy bands calculated in the previous sections, we now investigate the properties of the indirect excitons.
Flattening the electron and hole bands directly tunes the properties of the interlayer excitons and their strongly-correlated phases.
The exciton Rydberg energy $Ry^*= e^2/(4 \pi \epsilon 2 a_B^*)$, Bohr radius $a_B^*= \hbar^2 4 \pi \epsilon/(\mu e^2)$, and binding energy $E_b$ all depend on the effective reduced mass $\mu$, which is increased when the bands flatten.
These quantities are a measure of the strength of the electron-hole attraction, indicating the degree of difficulty of exciton dissociation \cite{kamban2020} and the dissociation temperature $k_B T\propto E_{b}$. \cite{Butov2016} 
The binding energy also provides insight into the properties of the strongly correlated excitonic states, including the exciton superfluid. 
The Berezinskii-Kosterlitz-Thouless transition temperature for the superfluid is proportional to $E_b$. \cite{Fogler2014, Conti2020}

\begin{figure}[h]
\centering
\includegraphics[width=0.85\columnwidth]{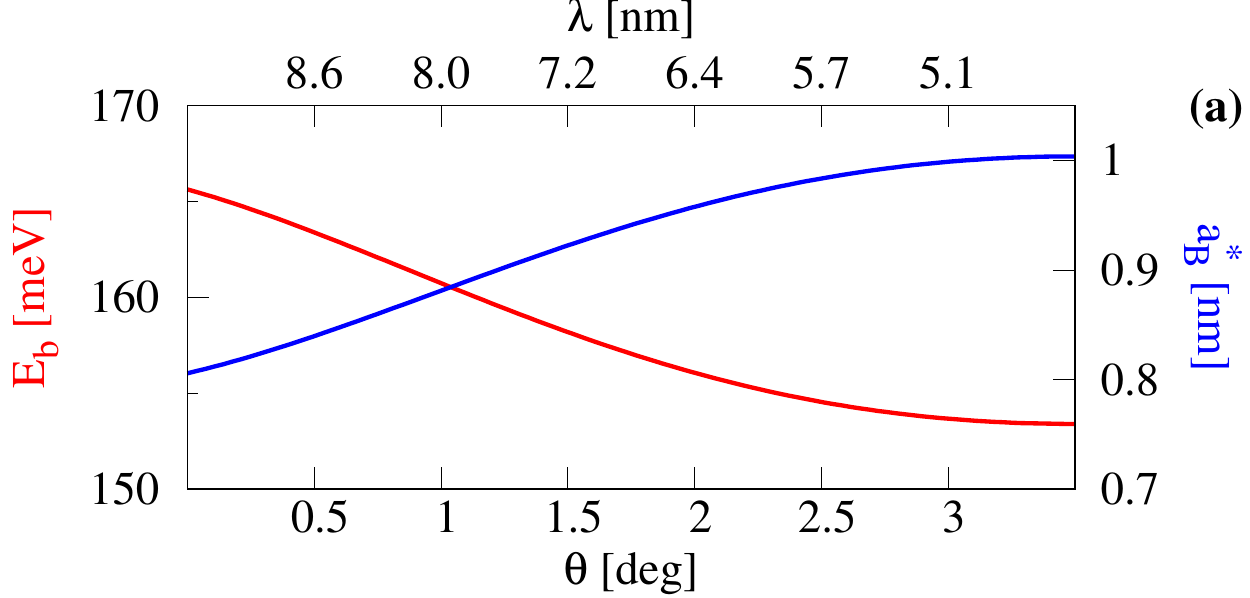}\\
\vspace{-0.08cm}
\includegraphics[width= 0.85\columnwidth]{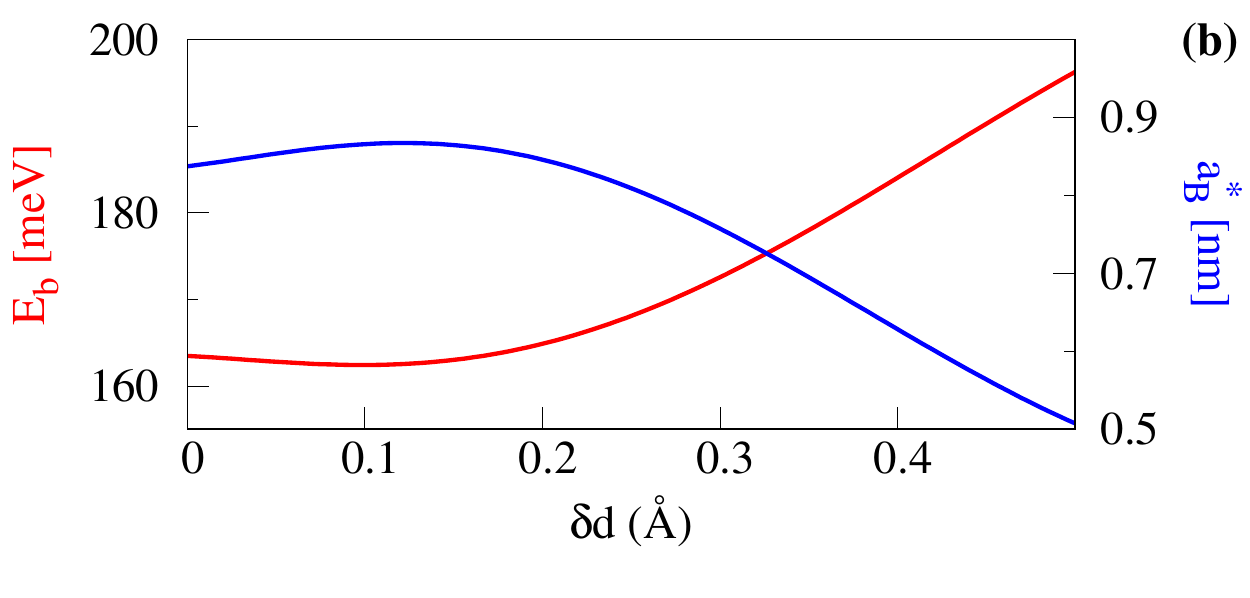}\\
\vspace{-0.30cm}
\includegraphics[width=0.85\columnwidth]{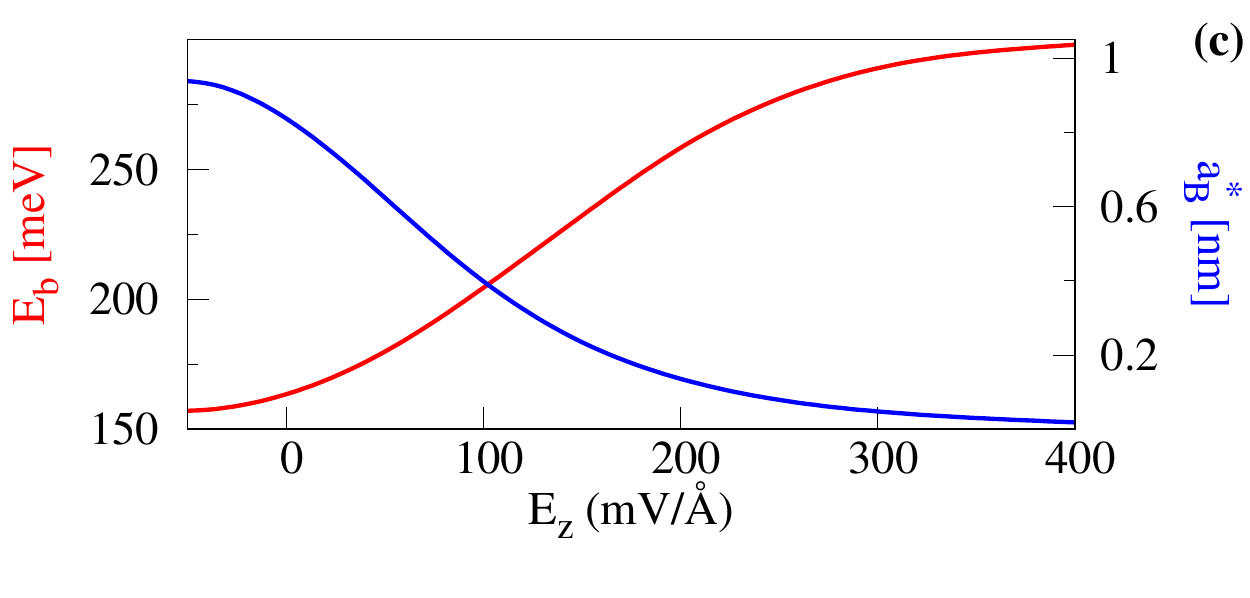}
\caption{Moir\'e interlayer exciton binding energy $E_b$ and Bohr radius $a_B^*$ in a MoS$_2$/WSe$_2$ vdWHB
(a) as a function of twist angle $\theta$ (and the corresponding period of the moir\'e potential $\lambda$), 
(b) as a function of reduction of the interlayer distance $\delta d$ upon pressure, 
and (c) as a function of a perpendicular electric field $E_z$.} 
\label{fig.moireExcitonEB}
\end{figure}

We use a variational approach solving eqn (\ref{eq.bindingenergy}) in Sec.~\ref{secEB} with the exciton wave-function modeled by an exponential.
Fig.~\ref{fig.moireExcitonEB} shows the resulting exciton binding energy $E_b$ and Bohr radius $a_B^*$ as a function of (a) twist angle, (b) interlayer distance tuned by vertical pressure, and (c) applied electric field.

Figure~\ref{fig.moireExcitonEB}a shows that with twisting, the moir\'e electron and hole masses have very limited tunability, so the exciton binding energy and Bohr radius do not greatly vary. 
For all the twist angles, $a_B^*$ remains an order of magnitude smaller than the moir\'e period $\lambda$ (top $x$-axis). 
This indicates that twisting has little effect on the localization of the excitons. In contrast, Fig.~\ref{fig.moireExcitonEB}b shows that if instead the interlayer distance is reduced by pressure, the binding energy increases by as much as $25\%$ and the Bohr radius decreases by $50\%$. 

The effect on the binding energy and Bohr radius is even more dramatic with application of a perpendicular electric field (Fig.~\ref{fig.moireExcitonEB}c).
The binding energy is enhanced by a factor of two for an increase in $E_z$ from $0$ to $400$ mV/\AA,  while the Bohr radius drops by a factor of $100$.
Here the moir\'e period $\lambda\sim 8.6$~nm for $\theta=0.5^\circ$, and 
the ratio of the effective Bohr radius to $\lambda$ decreases with increasing the electric field. 
By $E_z=400$ mV/\AA\, where the bands are ultra-flat, $a_B^*$ has decreased to two orders of magnitude less than $\lambda$.
These results indicate extreme localized exciton states and a striking evolution towards strong correlations.

\section{Conclusions}

In summary, we have determined the moir\'e bands for electrons and holes in MoS$_2$/WSe$_2$ vdWHB with small interlayer twist angles near $\theta \sim 0^\circ$, to tune under twisting, pressure, or electric fields the flatness of the electron and hole bands, for the purpose of investigating interlayer excitons in the strongly correlated regime. 

We have developed a continuum model parameterized from first principles, one that respects the crucial changes in the moir\'e potential from one stacking register to another.
The method is readily adaptable for other material combinations of van der Waals heterobilayers, as well as for small twist angles near $\theta \sim 60^\circ$.
The method is robust and reliable and does not suffer from the computational limitations in first principle calculations of moir\'e heterostructures, imposed by the large number of atoms in the unit cell.

We first demonstrate for this heterobilayer that in the vicinity of the $\Gamma$-point, the moir\'e potentials in the presence of a small interlayer twist \textit{do not} flatten the bands.  
This is opposite to the trend known for twisted bilayer graphene and expected for other homobilayers.
Although when the twist angle increases from zero the bands are seemingly flatter within the shorter Brillouin zone, we find that in fact, the overall effective mass of the quasi-particles remains practically unchanged. 

As an alternative idea for deepening the moir\'e potential, we  considered reducing the interlayer spacing by application of uniform  vertical pressure. This active manipulation was able to increase the effective mass of the hole by nearly an order of magnitude when the interlayer distance was decreased by $0.5$ \AA\ while, interestingly, the electron mass is left nearly unaltered.
Such different behavior of electrons and holes bears nontrivial consequences on the resulting properties of the interlayer excitons, of crucial importance to any further exotic strongly-correlated excitonic phases in van der Waals heterobilayers.

We find that as an even more effective strategy, applying an electric gating field perpendicular to the heterobilayer can dramatically deepen the moir\'e potential, thereby leading to strong increases of the moir\'e electron and hole effective masses. 
Concretely, in MoS$_2$/WSe$_2$ vdWHB with a $400$ mV/\AA\ electric field, the effective masses of the moir\'e electron and hole are both increased by a factor of $\sim 40$. 
This makes the moir\'e bands ultra-flat, with bandwidths as narrow as $0.05$ meV, the narrowest reported to date.

With such different yet complementary effects of these three manipulations, with their selective influences on electron and hole bands, and with the consequent tunability of the exciton binding energies and Bohr radii, we expect our results will help guide future works that seek strongly correlated electronic and excitonic phases in flat bands of 2D heterostructures.

\section{Theoretical Methods}

\subsection{Continuous Model}\label{secCM}
We build our continuum model for the moir\'e potential using parameters extracted from the DFT calculations. 
The $C_{3v}$ symmetry of the potential landscape is imposed by assuming a moir\'e potential expressed as
\begin{equation}\label{eq.V}
V_i(\vec{r}) = V_{i,1}|f_1(\vec{r})|^2 +  V_{i,2}|f_2(\vec{r})|^2,
\end{equation}
with the index $i = e (h)$ for electron (hole), 
$f_1(\vec{r}) = (e^{-i\vec{K}\cdot\vec{r}}+e^{-i\hat{C}_3\vec{K}\cdot\vec{r}}+e^{-i\hat{C}_3^2\vec{K}\cdot\vec{r}})/3$, and
$f_2(\vec{r}) = [e^{-i\vec{K}\cdot\vec{r}}+e^{-i(\hat{C}_3\vec{K}\cdot\vec{r}+\theta_s/2)}+e^{-i(\hat{C}_3^2\vec{K}\cdot\vec{r}+\theta_s})]/3$. 
The $\hat{C}_3$ operator represents a $120^\circ$ rotation, and $\theta_s = 4\pi/3$.

To determine the band structures, we represent the Hamiltonian in a finite difference scheme, incorporating the 2D moir\'e potential landscape of eqn (\ref{eq.V}) and assuming periodic boundary conditions with a Bloch wave approach.
The time independent Schr\"odinger equation is numerically solved separately for electrons and holes within the effective mass approximation.

\subsection{Density Functional Theory calculations}\label{secDFT}

The Density Functional Theory (DFT) calculations from first principles were performed using the Projector Augmented Wave (PAW) method \cite{blochl1994} implemented in the Vienna \emph{Ab-initio} Simulation Package (VASP). \cite{kresse1996,kresse1999} 
The generalized gradient approximation (GGA) from Perdew-Burke-Ernzerhof (PBE) is used for the exchange-correlation functional. \cite{perdew1996} 

We present details of DFT for untwisted MoS$_2$/WSe$_2$ vdWHB.
The van der Waals interactions between the MoS$_2$ and WSe$_2$ monolayers were included by the dispersion-corrected density functional (DFT-D3) method.\cite{grimme2010}
A vacuum spacing of $18$ \AA\ is employed along the out-of-plane direction to model an isolated heterostructure. 
To limit the induced strain, an average of the experimentally measured lattice constants of bulk MoS$_2$ ($3.160$ \AA)~\cite{takahashi1985} and bulk WSe$_2$ ($3.282\ $\AA)~\cite{schutte1987} is used as the in-plane lattice constant of the MoS$_2$/WSe$_2$ heterobilayer ($3.221$ \AA) for all the stackings considered.

Structural relaxation was performed using the conjugate-gradient method until the absolute value of the components of forces on out-of-plane positions converged to within $0.005$ eV/\AA. 
During the structural relaxation across all the stackings, the in-plane atomic positions were kept fixed.
Further refinement of the model to incorporate lattice relaxation in a heterostructure is left as an outlook, but should not affect the main conclusions of the work.
An energy cutoff of $450$ eV, energy convergence threshold of $10^{-7}$ eV, and $\Gamma$-centered k-mesh of $15\times15\times1$ were used for structural relaxation and self-consistent calculations.

Six valence electrons were used in the PAW pseudo-potential, $d^{5}s^{1}$ for W/Mo and $s^{2}p^{4}$ for S/Se.
Spin-orbit coupling was taken into account in all calculations except in structure relaxation. 
For each stacking, the MoS$_2$ and WSe$_2$ band edges are calculated at the $K$-point. 
\begin{figure}[!h]
\centering
\includegraphics[width=0.7\columnwidth]{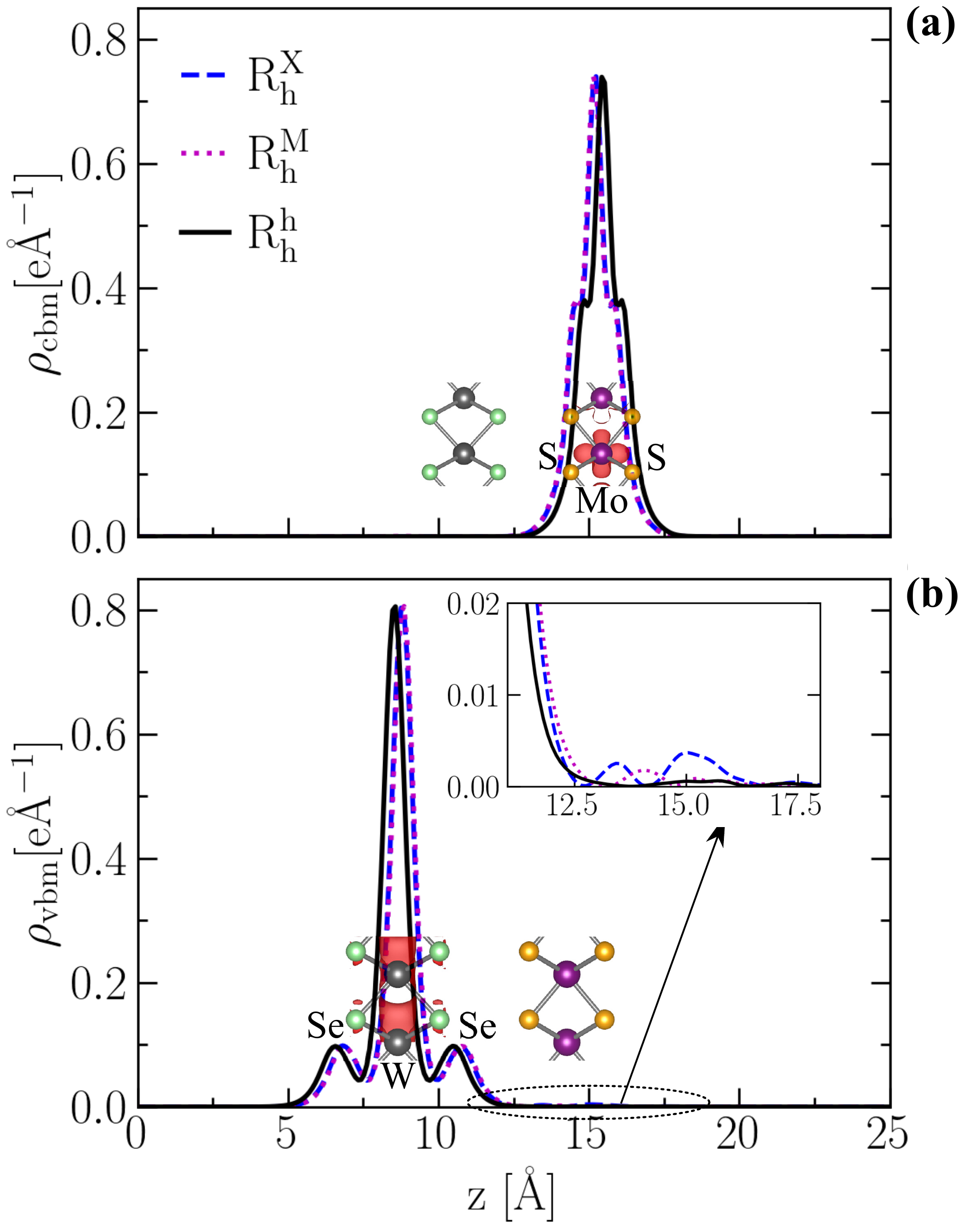}
\caption{Wave-function of (a) conduction band minimum and (b) valence band maximum states at the $K$-point, averaged along the perpendicular direction $z$, for untwisted MoS$_2$/WSe$_2$ vdWHB. 
The isosurface plots shown in red are for a value $0.004$ e/\AA$^3$ at $R^{h}_{h}$ stacking. 
Inset in panel (b) is a zoom-in showing the interlayer mixing. }
\label{fig.bandwf}
\end{figure}
\begin{figure}[!ht]
\centering
\includegraphics[width =0.65\columnwidth]{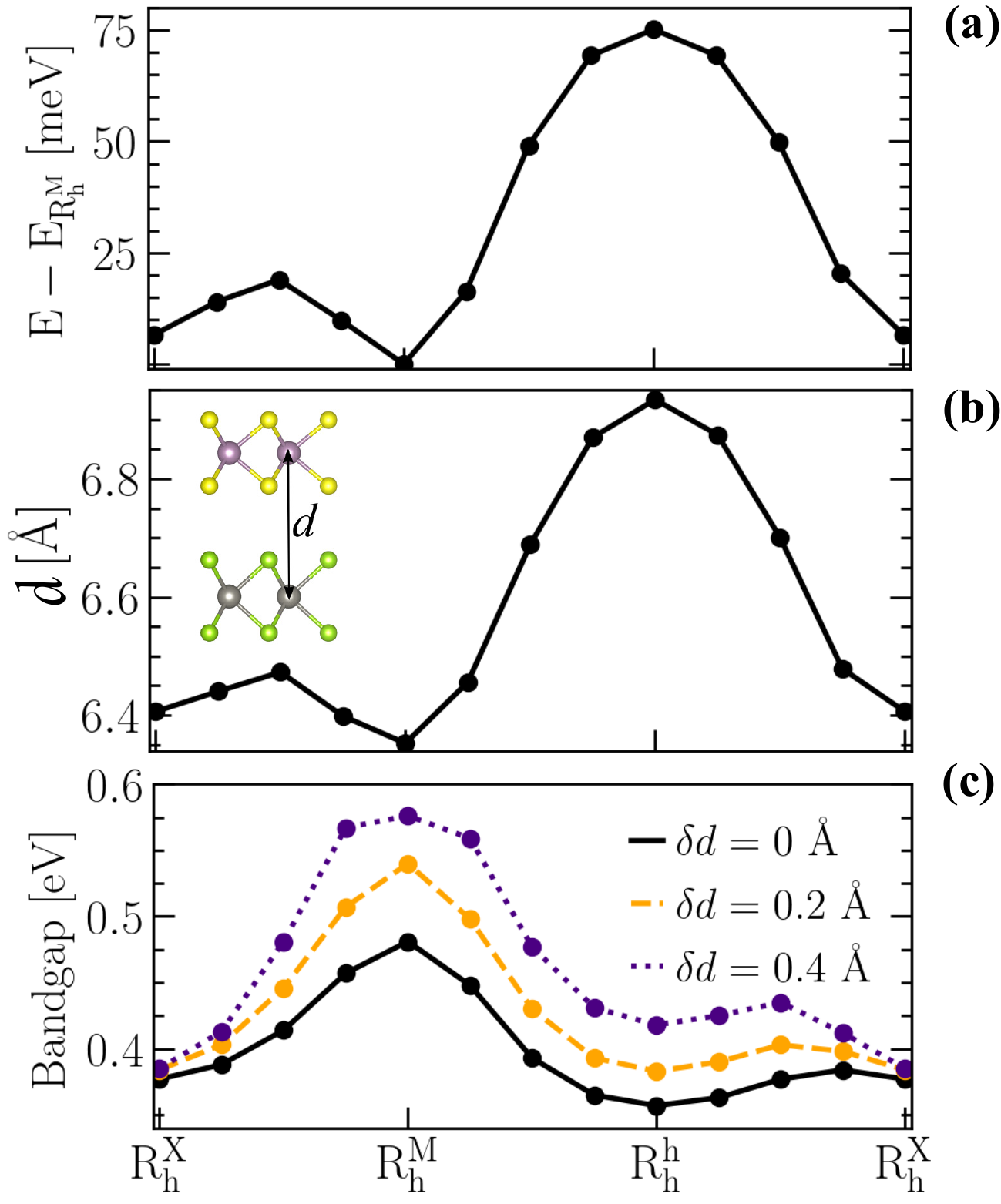}
\caption{(a) Variation of the total energy, (b) interlayer distance $d$, and (c) bandgap for the different stackings of aligned MoS$_2$/WSe$_2$ vdWHB. In panel (a) the energy is relative to the energetically favorable stacking (R$_{h}^{M}$).
In panel (c) $\delta d$ is the reduction in the interlayer distance.}
\label{energy_all}
\end{figure}

Figure~\ref{fig.bandwf} shows the wave-functions at the $K$-point of the conduction band minimum (CBM) and valence band maximum (VBM). 
The major contribution to the CBM charge density (panel (a)) comes from the $d$-orbital of the metal atom and is fully confined to the MoS$_2$ layer.
The VBM charge density (panel (b)) has contributions both from the metal and chalcogenide atoms, and it extends out from the WSe$_2$ layer to give rise to a small but finite interlayer mixing (see inset).

Figure~\ref{energy_all} shows for the different stackings, the variation in the total energy, the interlayer distances $d$, and the bandgap. 
We take $d$ as the distance between Mo and W atoms (see inset).
Fig.~\ref{energy_all}a shows that among all the sliding geometries explored, R$_{h}^{M}$ stacking is energetically the most favorable.
Fig.~\ref{energy_all}b shows that the interlayer distance has a sensitivity to the stacking order of up to $0.6$ \AA.
Uniform vertical pressure was modeled by reducing the interlayer distance by the same fixed amount across all the stackings without relaxation. 
Fig.~\ref{energy_all}c shows the evolution of the bandgap for the different stackings when the interlayer distance is reduced by $\delta d$.

\subsection{Exciton binding energy in moir\'e potential}\label{secEB}
A Wannier-Mott exciton in the presence of the moir\'e potential is described within an effective mass approximation by a Hamiltonian that includes (i) separate kinetic energy terms, (ii) moir\'e potentials for each electron and hole ($V_e$ and $V_h$, respectively), and (iii) an electron-hole interaction term $V_{eh}$: 
\begin{eqnarray}
H_{ex} = -\frac{\hbar^2}{2 M}\nabla^2_R -\frac{\hbar^2}{2 \mu}\nabla^2_r + V_e(R+\frac{m_h}{M}r) 
 + V_h(R-\frac{m_e}{M}r) + V_{eh}(r),
 \label{eq.Heh}
\end{eqnarray}
where $r$ is the in-plane relative coordinate and $R$ the center-of-mass coordinate of the electron-hole pair, and $M = m_e + m_h$ the total mass.

Unlike the system considered in Ref.~\citet{yu2017}, in which the energy scale of the electron-hole binding is much larger than that of the moir\'e potential landscape, here the moir\'e potentials reach hundreds of meV and are of the same order of magnitude as the electron-hole interaction.
With a deep moir\'e potential landscape, we assume that the moir\'e electron and hole band structures in isolated layers can be used to rewrite eqn (\ref{eq.Heh}) independent of $R$.
The binding energy $E_b$ of interlayer excitons in moir\'e potential is then obtained from  
\begin{equation}
 \left[-\frac{\hbar^2}{2 \mu^\prime}\nabla^2_r + V_{eh}(r)\right]\psi(r) = E_b\psi(r)\, . 
 \label{eq.bindingenergy}
\end{equation}
The modified reduced mass $\mu^\prime$ is determined from the moir\'e effective electron and hole masses. 
$V_{eh}(r)$ is the Keldysh potential for electrons and holes in different layers, \cite{brunetti2018}

\begin{equation}
V_{eh}(r)=\frac{e^2}{4 \pi  \epsilon  \epsilon_0} 
\frac{\pi}{2 r_0}
\left[H_0 \left(\frac{\sqrt{d^2+r^2}}{r_0}\right)-Y_0 \left(\frac{\sqrt{d^2+r^2}}{r_0}\right)\right],
\end{equation}
where $H_0$ and $Y_0$ are respectively the Struve and Bessel functions of the second kind, and $r_0 = 2\pi\chi/\epsilon$ is the screening length.
$\chi\approx 7$ nm is the 2D polarizability of the medium, and
$\epsilon\approx 4$ the dielectric constant for a MoS$_2$/WSe$_2$ vdWHB embedded in hexagonal boron nitride. \cite{Laturia2018}
The exciton occupies the energetically most favorable registry stacking.
In any case the dependence of $E_b$ on the local register due to varying interlayer distances at the different registries, is expected to be weak because the associated variation in $d$ is a tiny fraction of the exciton Bohr radius (Fig.~\ref{energy_all}b).

\section*{Author Contributions}
Conceptualization: S.C., A.C., D.N., and M.V.M;
Data curation: S.C.;
Formal Analysis: S.C., A.C., and T.P.;
Funding acquisition: S.C., A.C.;
Investigation: S.C., A.C, L.C., D.N., and M.V.M.;
Methodology: S.C., A.C., L.C., F.M.P., and M.V.M.;
Project administration: M.V.M.;
Resources: S.C., A.C., and T.P.;
Software: S.C., A.C., and T. P.;
Supervision: F.M.P., D.N., and M.V.M.;
Validation: F.M.P., D.N., and M.V.M.;
Visualization: S.C., and T.P.;
Writing – original draft: S.C., A.C., and T.P.;
Writing – review \& editing: S.C., A.C., T.P., L.C., F.M.P., D.N., and M.V.M.

\section*{Conflicts of interest}
The authors declare no conflict of interest.

\section*{Acknowledgements}
Discussions with Andrea Perali are gratefully acknowledged. S.C. and T.P. are supported by postdoctoral fellowships of the Research Foundation - Flanders (FWO-Vl).
A.C. and F.P. are supported by the Brazilian Council for Research (CNPq), through the PRONEX/FUNCAP, Universal, and PQ programs. The computational resources and services used in this work were provided by the VSC (Flem-
ish Supercomputer Center), funded by FWO and the
Flemish Government department EWI.


\balance


\providecommand*{\mcitethebibliography}{\thebibliography}
\csname @ifundefined\endcsname{endmcitethebibliography}
{\let\endmcitethebibliography\endthebibliography}{}

\end{document}